\documentclass[10pt,final,doublecolumn]{IEEEtran}
\usepackage{amsthm}
\usepackage{amsmath}
\usepackage{latexsym}
\usepackage{graphicx}
\usepackage{bbding}
\usepackage{indentfirst}
\usepackage{algorithm,algorithmic}
\usepackage{setspace}
\usepackage{float}
\usepackage{epstopdf}
\usepackage{amssymb}
\usepackage{amsfonts}
\usepackage{enumerate}
\usepackage{multicol}
\usepackage{color}
\usepackage{diagbox}
\usepackage{bm}
\usepackage{amssymb}
\usepackage{stfloats}
\usepackage{epstopdf}
\usepackage{threeparttable}
\usepackage{epstopdf}
\usepackage{threeparttable}
\usepackage{makecell}
\usepackage{cite}
\usepackage{graphicx}
\usepackage{subcaption}
\usepackage[labelsep=period]{caption} 
\usepackage[dvipsnames,table]{xcolor} 
\definecolor{mygray}{gray}{0.55}  
\usepackage{booktabs}       
\usepackage{siunitx}        

\IEEEoverridecommandlockouts
\allowdisplaybreaks[4]

\def\BibTeX{{\rm B\kern-.05em{\sc i\kern-.025em b}\kern-.08em
		T\kern-.1667eb\lower.7ex\hbox{E}\kern-.125emX}}
\usepackage{balance}

\begin{document}
	\title{Coupler Position Optimization and Channel Estimation for Flexible Coupler Antenna Aided Multiuser Communication}
\author{{Xiaodan Shao, \IEEEmembership{Member,~IEEE}, Chuangye Shan, Weihua Zhuang, \IEEEmembership{Fellow, IEEE}, Xuemin (Sherman) Shen, \IEEEmembership{Fellow, IEEE}}
		\thanks{X. Shao, W. Zhuang, and X. Shen are with the Department of Electrical and Computer Engineering, University of Waterloo, Waterloo, ON N2L 3G1, Canada (E-mail: \{x6shao, wzhuang, sshen\}@uwaterloo.ca)}
 }
\maketitle
	


\begin{abstract}
In this paper, we propose a distributed flexible
coupler antenna (FCA) array to enhance communication performance with low hardware cost. At each FCA, there is one fixed-position active antenna and
multiple passive couplers that can move within a designated
region around the active antenna. Moreover, each FCA
is equipped with a local processing unit (LPU). All LPUs
exchange signals with a central processing unit (CPU) for joint
signal processing.  We study an FCA-aided multiuser multiple-input multiple-output (MIMO) system, where an FCA array base station (BS)
is deployed to enhance the downlink communication between
the BS and multiple single-antenna users. We formulate optimization problems to maximize the achievable sum
rate of users by jointly optimizing the coupler positions and
digital beamforming, subject to movement constraints on the
coupler positions and the transmit power constraint. To address
the resulting nonconvex optimization problem, the digital beamforming is expressed as a function of the coupler position vectors,
which are then optimized using the proposed distributed coupler
position optimization algorithm. Considering a structured time domain pattern of pilots and coupler positions, pilot-assisted
centralized and distributed channel estimation algorithms are
designed under the FCA array architecture. Simulation results
demonstrate that the distributed FCA array achieves substantial rate
gains over conventional benchmarks in multiuser systems without
moving active antennas, and approaches the performance of fully
active arrays while significantly reducing hardware cost and
power consumption. Moreover, the proposed centralized channel
estimation algorithm outperforms the benchmark schemes in terms of both
pilot overhead and channel reconstruction accuracy. It is verified that reconfiguring passive coupler positions offers a cost-effective way to enhance wireless communication performance without moving active antennas.
\end{abstract}


\section{Introduction}
Multiple-input multiple-output (MIMO) is an important enabling technique for sixth-generation (6G) and beyond systems, since it can serve multiple mobile users on the same resource block concurrently \cite{exl, 8901196, kumar2025distributed, 10858129,10054381,10287775}. For instance, extremely large-scale MIMO can achieve significantly higher achievable rates by using more antennas than existing massive MIMO within a given space \cite{exl}. Passive MIMO, also known as intelligent reflecting surfaces (IRS)/reconfigurable intelligent surfaces (RIS), can enhance wireless network capacity in a cost-effective manner \cite{9867922,9133130,9724202,10893715,10443321,10540249,10555049}. An appealing property of MIMO is that its performance gains scale with the number of antennas.
Traditional MIMO systems, however, face two main challenges. First, higher data rates often require more antennas and thus larger installation space. Moreover, conventional MIMO typically requires a dedicated radio-frequency (RF) chain for each antenna, which increases hardware cost, power consumption, and transceiver size, and makes MIMO systems costly and bulky \cite{11162419,9791349}. Second, antennas in conventional MIMO systems are deployed at fixed positions, which limits the ability to adaptively reconfigure or optimize the wireless channel according to user distribution and environmental variations \cite{9903389}. 

On the other hand, most existing works on MIMO systems assume a centralized processing architecture, where a central processing unit (CPU) is connected to all antennas \cite{10858129}. In such systems, the received signals must be collected at the CPU to execute key tasks such as signal precoding and detection as well as channel estimation. As the number of antennas increases, centralized processing leads to significantly higher computational cost and complexity, together with longer processing delay, which may hinder the practical deployment of large-scale antenna arrays in future wireless networks. Moreover, centralized algorithms typically incur large communication cost between the antennas and the baseband processor, and they also require high computational cost due to large-scale signal processing \cite{yanqing1,8870236,4305460,7100912}. These issues call for a powerful and expensive CPU, which may become impractical or unaffordable when the antenna array size is large.

\begin{figure}[t!]
	\centering
	\setlength{\abovecaptionskip}{0.cm}
	\includegraphics[width=3.60in]{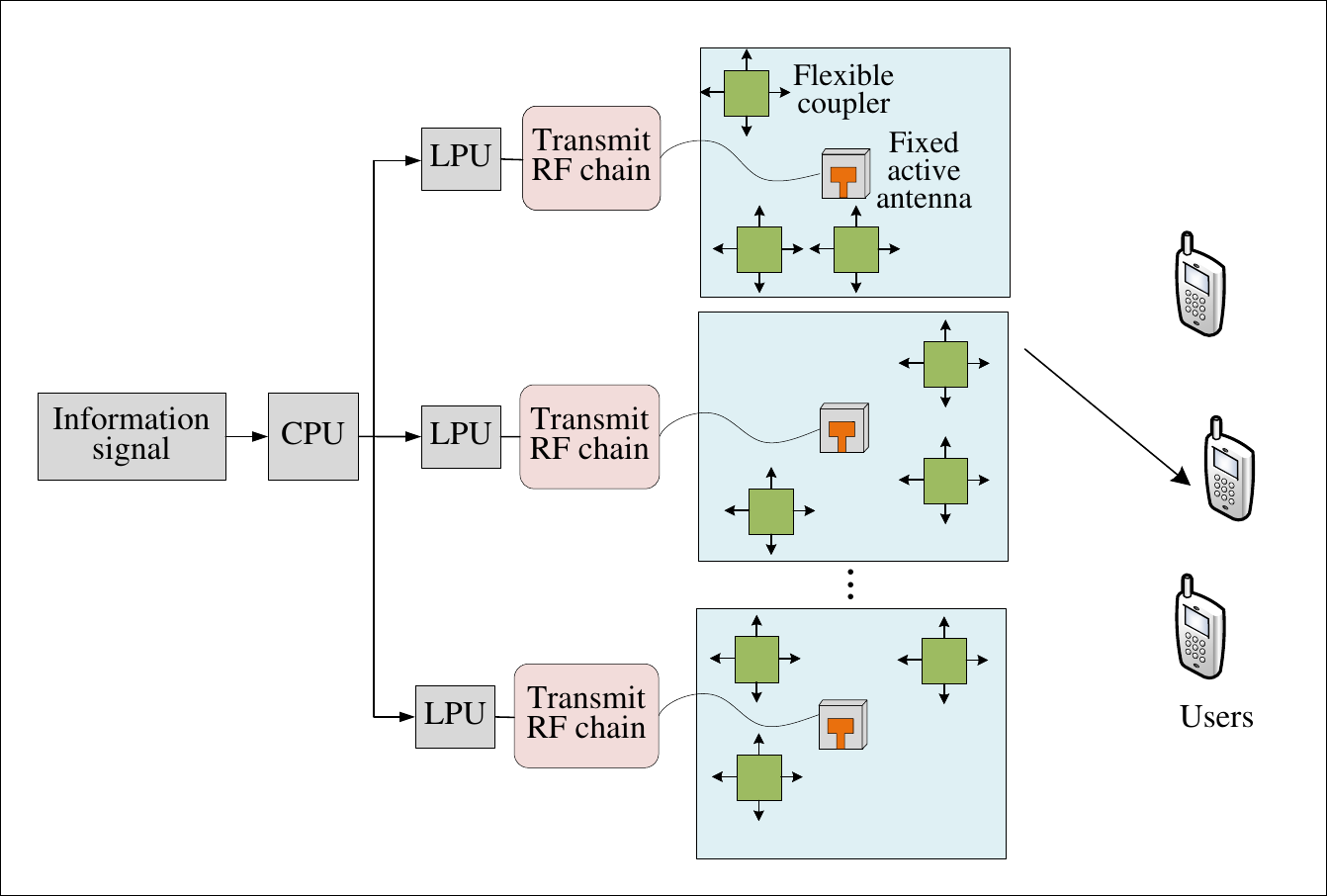}
	\caption{Proposed distributed FCA array for multiuser systems.}
	\label{FCAM}
			\vspace{-0.3cm}
\end{figure}

To overcome the above limitations, we propose in this paper a new distributed flexible
coupler antenna (FCA) array, as illustrated in Fig.~\ref{FCAM}, where each FCA is equipped with a local processing unit (LPU) to enable decentralized and parallel processing, thereby alleviating the computational burden at the CPU while improving communication performance. Moreover, by dispensing with the requirement that each antenna be connected to a dedicated RF chain, the FCA array consists of multiple FCAs, each comprising a single active antenna connected to one RF chain and multiple passive couplers. The passive couplers can be independently adjusted in position around their associated active antenna, while the active antenna remains fixed at the transceiver. By exploiting coupler movement and the mutual coupling between neighboring antenna and coupler elements, the passive couplers radiate through excitation induced by their associated active antenna. In the proposed distributed FCA array, 
all LPUs operate in a decentralized and parallel manner and exchange signals with the CPU for joint signal processing, such as channel estimation, precoding, and combining. The proposed architecture leverages both the CPU and the LPUs to not only offload computational tasks from the CPU but also reduce the frequency of data exchange between the distributed FCAs and the CPU by exploiting local computation and baseband signal processing at the LPUs.
As a result, the overall processing cost and latency are effectively reduced. 

Different from six-dimensional movable antennas (6DMA) \cite{shao20246d,6dma_dis,6DMA_JSTSP,11142311}, which enhance communication performance by translating and/or rotating active antennas together with their RF chains over relatively large regions\cite{11083675,10989638,wen,liu2024uav,UAV6DMA,11314850,10945745,near,jiang2025statistical,passive6DMA,free6DMA, wang20256d,shen20256d,IPA,6dmasensing}, in an FCA system, only the passive couplers are repositioned within a small local region where electromagnetic coupling is strong, while the active antennas and their RF chains remain fixed. This design substantially reduces transmit power consumption and mechanical control complexity, and enables faster and more practical position reconfiguration. Thus, the distributed FCA array features advantages such as a compact structure, low profile, and lightweight, which make it suitable for small-sized terminals and practical deployment.


The main contributions of this paper are summarized as follows. 
\begin{itemize}
	\item 
We propose a distributed FCA array to enhance communication performance and reduce centralized processing complexity by equipping each FCA with a local LPU. The FCA array consists of multiple FCAs, each containing one fixed-position active antenna and multiple passive couplers that can move within a designated region around each active antenna to enhance the multiuser communication. The distributed FCA array features a compact structure without moving the active antennas and their RF chains, making it particularly attractive for large-scale antenna deployments in small-sized devices.

	\item  
	Based on the distributed FCA array, we formulate an optimization problem to maximize the achievable sum rate of users by determining the positions of the passive couplers and the transmit beamforming, subject to their practical movement and transmit power constraints. To solve this nonconvex problem, we develop a successive convex approximation (SCA)-based distributed coupler-position optimization algorithm, where the LPUs update the coupler positions in parallel via local projections and the CPU broadcasts the required gradient information and updates the minimum mean square error (MMSE) precoder.

	\item 
By exploiting a structured training protocol with multiple pilot blocks and blockwise fixed coupler positions, we develop two pilot-assisted channel estimation algorithms under the distributed FCA architecture, namely, centralized scheme and distributed scheme. In the centralized scheme, the CPU stacks pilot-correlated observations across all FCAs and performs sparse recovery to estimate the angular support and path gains, enabling channel reconstruction for arbitrary coupler positions. In the distributed scheme, each LPU conveys low-overhead support information to the CPU, which carries out lightweight fusion and path gain estimation using low-dimensional statistics.
	
	\item
	Extensive simulations validate the effectiveness of the proposed distributed FCA array and the associated optimization and channel estimation designs. The results show that the proposed FCA array can closely approach the achievable rate of a fully active array while using significantly fewer active antennas, due to the interference mitigation, mechanical beamforming gain, and fading mitigation enabled by coupler repositioning. Moreover, simulations demonstrate the efficiency of the proposed channel estimation schemes.
\end{itemize}

The remainder of this paper is organized as follows. Section~II describes the system model, including the distributed FCA array structure, the channel model, mechanical beamforming, and the signal model of the distributed FCA array. Section~III formulates the optimization problem for maximizing the achievable sum rate through coupler position design and then presents the proposed distributed coupler position optimization algorithm. Section~IV presents the proposed centralized and distributed channel estimation algorithms for the designed FCA array. Section~V provides numerical results and corresponding discussions. Finally, Section~VI concludes this work.

\emph{Notations.}
Boldface uppercase and lowercase letters denote matrices and vectors, respectively,
$(\cdot)^{*}$, $(\cdot)^{H}$, and $(\cdot)^{T}$ denote complex conjugation, conjugate transpose, and transpose, respectively.
For scalar $a$, $|a|$ denotes its magnitude.
For vector $\mathbf{a}$, $\|\mathbf{a}\|_2$ denotes the $\ell_{2}$ norm. For matrix $\mathbf A$, $\|\mathbf A\|_{F}$ denotes the Frobenius norm.
Operator $\mathrm{diag}(\mathbf{x})$ denotes a diagonal matrix whose diagonal is $\mathbf{x}$,
$[\mathbf{a}]_{j}$ denotes the $j$-th entry of vector $\mathbf{a}$, $\mathbb{E}[\cdot]$ denotes statistical expectation, $[\mathbf{A}]_{i,j}$ denotes the $(i,j)$-th entry of a matrix $\mathbf{A}$, $[\mathbf A]_{:,k}$ and $[\mathbf A]_{k,:}$ denote the $k$-th column and the $k$-th row of $\mathbf A$, respectively,
$\mathbf{I}_{M}$ denotes the $M\times M$ identity matrix,
$\mathrm{blkdiag}\{\mathbf{A}_{1},\ldots,\mathbf{A}_{M}\}$ denotes a block-diagonal matrix with diagonal sub-matrices given by $\mathbf{A}_{1},\ldots,\mathbf{A}_{M}$,
$\mathbb{R}$ and $\mathbb{C}$ denote the real and complex fields, respectively,
$\mathcal{CN}(0,\sigma^{2})$ denotes a complex Gaussian distribution with mean $0$ and variance $\sigma^{2}$,
$\Re\{\cdot\}$ denotes the real part of a complex quantity,
 $|\mathcal I|_{\mathrm c}$ denotes the cardinality of index set $\mathcal I$, $\mathrm{supp}(\mathbf x)$ denotes the support set of vector $\mathbf x$, $\mathrm{tr}(\cdot)$ denotes the trace of a square matrix, and $\nabla f(\mathbf x)$ denotes the gradient of function $f$ with respect to $\mathbf x$.

\section{System Model}
\subsection{Distributed Flexible Coupler Array}
As shown in Fig.~\ref{FCAM}, the proposed FCA array consists of $M$ FCAs. Each FCA comprises one fixed-position active antenna and $N$ passive couplers, which can move within a designated region, i.e., $\mathcal{C}_m$ surrounding the $m$-th active antenna. Each active antenna is connected to a single RF chain. The positions of passive couplers are mechanically adjustable with the aid of actuation components such as micro-electro-mechanical systems (MEMS), which offer low power consumption and fast response.
Moreover, each FCA is equipped with local computing hardware, i.e., an LPU, which carries out the necessary baseband processing tasks, such as channel estimation and signal precoding and combining, in a decentralized and parallel fashion, as shown in Fig.~\ref{FCAM}. Each LPU is connected to the common CPU at the base station (BS) via separate flexible cables, such as coaxial cables. These cables supply power to the FCA and also facilitate control and signal exchange between the LPU and the CPU. The users are each assumed to be equipped with a single fixed-position antenna. In this architecture, the signals radiated by the passive couplers are generated through induced currents excited by their associated active antenna via near-field electromagnetic coupling, without requiring additional RF chains or feeding networks. By reconfiguring only the passive coupler positions within a small local region, the proposed FCA array enhances wireless communication performance while avoiding the movement of active antennas or RF chains, thereby significantly reducing mechanical control complexity and energy consumption.

\subsection{Channel Model} 
As illustrated in Fig.~\ref{FCAM}, we consider a downlink multiuser wireless communication system where an FCA array is deployed at the BS.  
The BS serves $K$ single-antenna users.  
The FCA array can be reconfigured by mechanically adjusting the positions of its passive couplers.

We adopt a two-dimensional Cartesian coordinate system. 
The transmit array lies in the $x$–$y$ plane. 
The $M$ active antennas form a uniform linear array along the $y$-axis. 
The first active antenna is located at origin $\mathbf q_1 = [0,0]^{\mathrm T}$ and the $m$-th active antenna is located at
\begin{align}\label{gt3}
	\mathbf q_m = [0,(m-1)d_y]^{\mathrm T}, \quad m \in \mathcal{M},
\end{align}
where $\mathcal{M}=\{1, 2, \ldots,M\}$ and $d_y$ denotes the spacing between adjacent active antennas.

Let $\mathcal{K}=\{1,2,\ldots,K\}$ denote the set of user indices.  
For user $k\in\mathcal{K}$, we consider a narrowband geometric multipath channel with $L_k$ paths.  
The azimuth angle of departure of the $\ell$-th path from the BS  is denoted by $\phi_{k,\ell}$ and the corresponding complex path gain is denoted by $\alpha_{k,\ell}$, $\ell=1,2,\ldots,L_k$.  
Let $\lambda$ denote the carrier wavelength.  
The steering vector of the active-antenna array for angle $\phi_{k,\ell}$ is
\begin{align}\label{ay_def}
	\mathbf a_{y}(\phi_{k,\ell})
	=& \big[
	1,\,
	e^{-j\frac{2\pi}{\lambda} d_y \sin\phi_{k,\ell}},\,
	\ldots,
	e^{-j\frac{2\pi}{\lambda} (M-1) d_y \sin\phi_{k,\ell}}
	\big]^{\mathrm T}. 
\end{align}
The channel vector between all active antennas and user $k$ is
\begin{align} \label{part1_mu}
	\mathbf h_{\mathrm A,k}
	&= 
	\sum_{\ell=1}^{L_k} \alpha_{k,\ell}\,
	\mathbf a_{y}(\phi_{k,\ell})
	\in \mathbb C^{M\times 1}. 
\end{align}

The two-dimensional position of the $n$-th coupler associated with the $m$-th active antenna is
\begin{align}\label{gt6}
	\mathbf p_{n,m} = [x_{n,m},\, y_{n,m}]^{\mathrm T}\in\mathbb R^{2}, 
	\quad n \in \mathcal{N},\; m \in \mathcal{M},
\end{align}
where $\mathcal{N}=\{1,2,\ldots,N\}$. 
We collect all coupler positions into vector 
\begin{align}
	\mathbf p
	=
	\big[
	\mathbf p_{1}^{\mathrm T},\mathbf p_{2}^{\mathrm T},
	\ldots,\mathbf p_{M}^{\mathrm T}
	\big]^{\mathrm T},
\end{align}
where $\mathbf p_m=[\mathbf p_{1,m}^{\mathrm T}, \mathbf p_{2,m}^{\mathrm T},\ldots,\mathbf p_{N,m}^{\mathrm T}]^{\mathrm T}\in\mathbb R^{2N}$ is the position vector of the couplers in the $m$-th FCA. 

For user $k$ and its $\ell$-th path, we denote the two-dimensional direction vector in the $x$–$y$ plane as
\begin{align}
	\boldsymbol\kappa_{k,\ell}
	=
	[\cos\phi_{k,\ell},\, \sin\phi_{k,\ell}]^{\mathrm T}\in\mathbb R^{2}.
\end{align}
The coupler steering vector for user $k$ on path $\ell$ is given by
\begin{align}
	\mathbf a_{\mathrm C,k,\ell}(\mathbf p)
	=
	\big[
	\mathbf a_{\mathrm C,k,\ell}(\mathbf p_1)^{\mathsf T},
	\mathbf a_{\mathrm C,k,\ell}(\mathbf p_2)^{\mathsf T},
	\ldots,
	\mathbf a_{\mathrm C,k,\ell}(\mathbf p_M)^{\mathsf T}
	\big]^{\mathsf T},
\end{align}
where $\mathbf a_{\mathrm C,k,\ell}(\mathbf p_m)$ denotes the coupler steering vector associated with user $k$, path $\ell$, and the $m$-th FCA, and is given by
\begin{align}
	\mathbf a_{\mathrm C,k,\ell}(\mathbf p_m)
	=
	\big[
	e^{-j \tfrac{2\pi}{\lambda}\boldsymbol\kappa_{k,\ell}^{\mathsf T}\mathbf p_{1,m}},
	\ldots,
	e^{-j \tfrac{2\pi}{\lambda}\boldsymbol\kappa_{k,\ell}^{\mathsf T}\mathbf p_{N,m}}
	\big]^{\mathsf T}.
\end{align}
The channel vector between all couplers and user $k$ is 
\begin{align}\label{part2_mu}
	\mathbf h_{\mathrm C,k}(\mathbf p)
	=
	\sum_{\ell=1}^{L_k} \alpha_{k,\ell}
	\mathbf a_{\mathrm C,k,\ell}(\mathbf p)\in \mathbb{C}^{MN \times 1}.
\end{align}

Based on \eqref{part1_mu} and \eqref{part2_mu}, the overall multipath channel vector from the FCA array to user $k$ is given by 
\begin{align} \label{vty_mu}
	\mathbf h_k(\mathbf p)
	=
	\big[
	\mathbf h_{\mathrm A,k}^{\mathrm T},
	\mathbf h_{\mathrm C,k}^{\mathrm T}(\mathbf p)
	\big]^{\mathrm T}\in \mathbb{C}^{(MN+M) \times 1},
\end{align}
which depends on coupler positions $\mathbf p$ while the active antenna positions are fixed.  
Collecting the channel vectors of all users yields the following multiuser MIMO channel matrix
\begin{align}
	\mathbf H(\mathbf p)
	=
	\big[
	\mathbf h_1(\mathbf p),\ldots,\mathbf h_K(\mathbf p)
	\big]\in\mathbb C^{(MN+M)\times K}.
\end{align}

\subsection{Mechanical Beamforming}
Let $\mathbf{i}_{\mathrm{A}} \in \mathbb{C}^{M \times 1}$ denote the current vector of all active antennas, and
$\mathbf{i}_{\mathrm{C}} \in \mathbb{C}^{M N \times 1}$ denote the current vector of all couplers. The current vector across all active antennas and couplers is given by
\begin{align}\label{ittx}
	\mathbf{i} = [\mathbf{i}_{\mathrm{A}}^{\mathrm{T}}, \mathbf{i}_{\mathrm{C}}^{\mathrm{T}}]^{\mathrm{T}}\in \mathbb{C}^{(MN+M) \times 1}.
\end{align}

Similarly, let the active-antenna voltage vector be $\mathbf{v}_{\mathrm{A}} \in \mathbb{C}^{M \times 1}$ and the coupler voltage vector be $\mathbf{v}_{\mathrm{C}} \in \mathbb{C}^{M N \times 1}$. The voltage vector across all active antennas and couplers is given by
\begin{align}\label{ittxv}
	\mathbf{v} = [\mathbf{v}_{\mathrm{A}}^{\mathrm{T}}, \mathbf{v}_{\mathrm{C}}^{\mathrm{T}}]^{\mathrm{T}}\in \mathbb{C}^{(MN+M) \times 1}.
\end{align}

In general, the mutual impedance matrix between all couplers and all  
active antennas can be denoted by $\bar{\mathbf Z}(\mathbf p)\in \mathbb C^{MN\times M}$ \cite{kalis2014parasitic}.
We denote the mutual coupling vector between the $j$-th
active antenna and the $N$ couplers associated with the $i$-th
active antenna by $\bar{\mathbf z}_{i,j}\in \mathbb C^{N\times 1}$. Then, $\bar{\mathbf Z}(\mathbf p)$ can be written as
\begin{align}
	\bar{\mathbf Z}(\mathbf p)
	&=
	\begin{bmatrix}
		\bar{\mathbf z}_{1,1} & \cdots & \bar{\mathbf z}_{1,M}\\
		\vdots               & \ddots & \vdots\\
		\bar{\mathbf z}_{M,1} & \cdots & \bar{\mathbf z}_{M,M}
	\end{bmatrix}\in \mathbb{C}^{MN \times M}.                                         \label{eq:Zbar_block}
\end{align}
Let $\widetilde{\mathbf Z}\in\mathbb C^{M\times M}$ denote the mutual
impedance among all active antennas and 
$\hat{\mathbf Z}(\mathbf p)\in \mathbb C^{MN\times MN}$ denote the mutual
impedance among all couplers. The overall mutual impedance
matrix of the FCA array is thus given by 
\begin{align}
	\mathbf Z(\mathbf p)
	&=
	\begin{bmatrix}
		\widetilde{\mathbf Z} & \bar{\mathbf Z}^{\mathrm T}(\mathbf p)\\
		\bar{\mathbf Z}(\mathbf p) & \hat{\mathbf Z}(\mathbf p)
	\end{bmatrix}
	\in \mathbb C^{(MN+M)\times (MN+M)}.   \label{eq:Z_full}
\end{align}

For practical implementation considerations, we assume that the active antennas are physically isolated by metallic plates. The center-to-center distance between any two active antennas is assumed to be significantly larger than the distance between each active antenna and its associated passive couplers, as well as the distances among the couplers themselves. Consequently, the mutual coupling between ports associated with different FCAs becomes negligible and can be ignored. Under this assumption, the overall mutual impedance matrix exhibits a block-diagonal structure. Accordingly, we define the mutual impedance sub-matrix corresponding to the $m$-th FCA as
\begin{align}
	\mathbf Z_m(\mathbf p_m)
	&=
	\begin{bmatrix}
		\tilde{z}_m &
		\overline{\mathbf Z}_m^{\mathrm T}(\mathbf p_m)\\[0.5mm]
		\overline{\mathbf z}_m(\mathbf p_m) &
		\hat{\mathbf Z}_m(\mathbf p_m)
	\end{bmatrix}\in \mathbb C^{(N+1)\times (N+1)},    \label{eq:Zm_def}
\end{align}
where $\tilde{z}_m\in\mathbb C$ denotes the self-impedance of the $m$-th active antenna, 
$\overline{\mathbf z}_m(\mathbf p_m)\in\mathbb C^{N\times 1}$  collects the mutual impedances between this active antenna and its associated $N$ couplers, and
$\hat{\mathbf Z}_m(\mathbf p_m)\in\mathbb C^{N\times N}$ denotes the mutual impedance matrix among these couplers. Under the above block-diagonal assumption,
the overall mutual impedance matrix can be simplified to 
\begin{align}
	\mathbf Z(\mathbf p)
	= \mathrm{blkdiag}\big\{
	\mathbf Z_1(\mathbf p_1),\ldots,\mathbf Z_M(\mathbf p_M)
	\big\}.                                            \label{eq:Z_blockdiag}
\end{align}

Let $X_{n,m}$ denote the load impedance at the port of the $n$-th coupler associated with the $m$-th active antenna. The diagonal load-impedance matrix connected to the $N$ couplers of the $m$-th FCA is given by \cite{kalis2014parasitic}
\begin{align}
		\mathbf X_m
	= \mathrm{diag}\!\big\{X_{1,m},\ldots,X_{N,m}\big\}
	\in\mathbb C^{N\times N}.                            \label{eq:Xm_def}
\end{align}

Let $s\in\mathbb C$ denote the transmit information symbol with zero mean and unit variance. Antenna current vector $\mathbf i$ is adopted as the equivalent transmit precoding (beamforming) vector, such that the radiated signal is given by $\mathbf i s$. For the $m$-th FCA, define $\mathbf v_{\mathrm C,m}\in\mathbb C^{N\times 1}$ as the port-voltage vector across the load impedances at the network side, and define the load-voltage drop along the direction of coupler current vector $\mathbf i_{\mathrm C,m}\in\mathbb C^{N\times 1}$. By Kirchhoff's voltage law \cite{lara2015kirchhoff}, we have
\begin{align}
	\mathbf v_{\mathrm C,m} = -\mathbf X_m \mathbf i_{\mathrm C,m}\, s,\quad m\in\mathcal M, \label{eq:vCm_load}
\end{align}
where $\mathcal M=\{1,2,\cdots,M\}$.
Moreover, the port voltages and currents of the linear
multi-port network obey the impedance relation \cite{5446312},
\begin{align}
	\mathbf v = \mathbf Z(\mathbf p)\mathbf i s.         \label{eq:vZis}
\end{align}
Partitioning \eqref{eq:vZis} according to
$\mathbf v = [\mathbf v_{\mathrm A}^{\mathrm T},\mathbf v_{\mathrm C}^{\mathrm T}]^{\mathrm T}$ and
$\mathbf i = [\mathbf i_{\mathrm A}^{\mathrm T},\mathbf i_{\mathrm C}^{\mathrm T}]^{\mathrm T}$ and using
\eqref{eq:Z_blockdiag} yield
\begin{align}
	\mathbf v_{\mathrm C,m}
	= \overline{\mathbf z}_m(\mathbf p_m) i_{\mathrm A,m}s
	+ \hat{\mathbf Z}_m(\mathbf p_m)\mathbf i_{\mathrm C,m}s,  ~m\in\mathcal M,
	\label{eq:vCm_network}
\end{align}
where $i_{\mathrm A,m}$ is the current at the $m$th active antenna, and
$\mathbf i_{\mathrm C,m}\in\mathbb C^{N\times 1}$ collects the currents of the associated $N$ couplers. Equating \eqref{eq:vCm_load} and
\eqref{eq:vCm_network} gives
\begin{align}
	\mathbf i_{\mathrm C,m}
	&= -\big(
	\hat{\mathbf Z}_m(\mathbf p_m)+\mathbf X_m
	\big)^{-1}
	\overline{\mathbf z}_m(\mathbf p_m) i_{\mathrm A,m}     \nonumber\\
	&= -\mathbf w_m(\mathbf p_m) i_{\mathrm A,m},      \label{eq:iCm_def}
\end{align}
where
\begin{align}
	\mathbf w_m(\mathbf p_m)
	=
	\big(
	\hat{\mathbf Z}_m(\mathbf p_m)+\mathbf X_m
	\big)^{-1}
	\overline{\mathbf z}_m(\mathbf p_m)
	\in\mathbb C^{N\times 1},
\end{align}
is the local precoding/beamforming vector of the $m$-th FCA
antenna. We refer to $\mathbf w_m(\mathbf p_m)$ as the mechanical beamforming vector, since its beamforming weights can be reconfigured through adjusting the coupler positions, $\mathbf p_m$. 
Stacking the mechanical beamforming vectors of all FCAs gives the following block-diagonal mechanical beamforming matrix,
\begin{align}
\!\!\!	\mathbf W(\mathbf p)
	= \mathrm{blkdiag}\big\{
	\mathbf w_1(\mathbf p_1),\ldots,\mathbf w_M(\mathbf p_M)
	\big\}
	\in\mathbb C^{MN\times M}.                           \label{eq:W_blockdiag}
\end{align}

\subsection{Signal Model}
For user $k\in\mathcal K=\{1,2,\cdots,K\}$, let $s_k\in\mathbb C$ denote the transmit information symbol with
$\mathbb E\{|s_k|^2\}=1$. We collect all user symbols into the vector,
\begin{align}
	\mathbf s = [s_1,\ldots,s_K]^{\mathsf T}\in\mathbb C^{K\times 1},
\end{align}
which satisfies $\mathbb E\{\mathbf s\mathbf s^{\mathsf H}\}=\mathbf I_K$.

For a given coupler-position vector $\mathbf p$, we use the active-antenna current vector as the equivalent digital beamforming vector across the $M$ FCAs. Specifically, for user $k\in\mathcal K$, we define
\begin{align}
	\mathbf i_{\mathrm A,k}
	=
	\big[i_{\mathrm A,1,k},\ldots,i_{\mathrm A,M,k}\big]^{\mathsf T}
	\in\mathbb C^{M\times 1},
\end{align}
where $i_{\mathrm A,m,k}$ is the current injected into the $m$-th active antenna for user $k$.
Stacking $\{\mathbf i_{\mathrm A,k}\}_{k=1}^{K}$ yields digital precoding matrix
\begin{align}
	\mathbf U
	=
	\big[\mathbf i_{\mathrm A,1},\ldots,\mathbf i_{\mathrm A,K}\big]
	\in\mathbb C^{M\times K}.                                  \label{eq:IA_def}
\end{align}

Next, we absorb user symbols $\{s_k\}_{k=1}^{K}$ into the injected currents and obtain the instantaneous active-antenna current vector as
\begin{align}
	\mathbf i_{\mathrm A}
	=
	\sum_{k=1}^{K}\mathbf i_{\mathrm A,k}s_k
	=
	\mathbf U\mathbf s
	\in\mathbb C^{M\times 1},                               \label{eq:iA_inst_mu}
\end{align}
which equivalently redefines the effective complex amplitude and phase of each symbol $s_k$. According to \eqref{eq:iCm_def}, the corresponding coupler current vector is given by
\begin{align}
	\mathbf i_{\mathrm C}
	=
	-\mathbf W(\mathbf p)\mathbf i_{\mathrm A}
	=
	-\mathbf W(\mathbf p)\mathbf U\mathbf s
	\in\mathbb C^{MN\times 1}.                                 \label{eq:iC_inst_mu}
\end{align}
Stacking the currents of all active antennas and all couplers yields the instantaneous port current vector
\begin{align}
	\mathbf i_{\mathrm{tot}}
	&=
	\begin{bmatrix}
		\mathbf i_{\mathrm A}\\
		\mathbf i_{\mathrm C}
	\end{bmatrix}
	=
	\begin{bmatrix}
		\mathbf I_M\\
		-\mathbf W(\mathbf p)
	\end{bmatrix}
	\mathbf U\mathbf s
	=
	\widetilde{\mathbf W}(\mathbf p)\mathbf U\mathbf s,
	\label{eq:i_tot_inst_mu}
\end{align}
where
\begin{align}
	\widetilde{\mathbf W}(\mathbf p)
	=
	\begin{bmatrix}
		\mathbf I_M\\
		-\mathbf W(\mathbf p)
	\end{bmatrix}
	\in\mathbb C^{(M+MN)\times M},                            \label{eq:Wtilde_def_mu}
\end{align}
denotes the extended mechanical beamforming matrix.

Then, the received signal at user $k$ is given by
\begin{subequations}
\begin{align}
	&y_k
	= \mathbf h_k^{\mathsf T}(\mathbf p)\,
	\mathbf i_{\mathrm{tot}} + n_k \label{eq:yk_mu1}\\
	&= \mathbf h_k^{\mathsf T}(\mathbf p)\,
	\widetilde{\mathbf W}(\mathbf p)\mathbf U\mathbf s + n_k \label{eq:yk_mu0}\\
	&= \mathbf h_k^{\mathsf T}(\mathbf p)\,
	\widetilde{\mathbf W}(\mathbf p)\mathbf i_{\mathrm A,k}s_k
	+ \sum_{\substack{j\in\mathcal K,\, j\neq k}}
	\mathbf h_k^{\mathsf T}(\mathbf p)\,
	\widetilde{\mathbf W}(\mathbf p)\mathbf i_{\mathrm A,j}s_j
	+ n_k,                                                   \label{eq:yk_mu}
\end{align}
\end{subequations}
where $n_k\sim\mathcal{CN}(0,\sigma_k^2)$ denotes receive complex additive Gaussian noise at user $k$ with zero mean and variance $\sigma_k^2$.

Thus, the signal-to-interference-plus-noise ratio (SINR) for decoding $s_k$ at user $k$ is given by
\begin{align}
	\gamma_k(\mathbf p,\mathbf U)
	=
	\frac{\big|\mathbf h_k^{\mathsf T}(\mathbf p)\,
		\widetilde{\mathbf W}(\mathbf p)\mathbf i_{\mathrm A,k}\big|^2}
	{\displaystyle
		\sum_{\substack{j\in\mathcal{K},\, j\neq k}}
		\big|
		\mathbf h_k^{\mathsf T}(\mathbf p)\,
		\widetilde{\mathbf W}(\mathbf p)\mathbf i_{\mathrm A,j}
		\big|^2
		+ \sigma_k^{2}}.                                      \label{eq:SINR_mu}
\end{align}
Accordingly, the achievable downlink sum rate per hertz is given by
\begin{align}
	R(\mathbf p,\mathbf U)
	=
	\sum_{k=1}^{K}
	\log_2\big(1+\gamma_k(\mathbf p,\mathbf U)\big).          \label{eq:sumrate_mu}
\end{align}

\section{Optimization of Coupler Positions}

\subsection{Problem Formulation}
According to multiport circuit theory \cite{5446312}, the average transmit power equals the expected real power 
\begin{align}
	P
	=
	\mathbb E\!\left\{
	\Re\big\{
	\mathbf i_{\mathrm{tot}}^{\mathsf H}\,
	\mathbf v_{\mathrm{tot}}
	\big\}
	\right\},
	\label{eq:P_def_start_mu}
\end{align}
where the expectation is taken over the random information symbols involved in $\mathbf i_{\mathrm{tot}}$.
Using the impedance relation $
\mathbf v_{\mathrm{tot}}
=
\mathbf Z(\mathbf p)\mathbf i_{\mathrm{tot}}$,
we obtain
\begin{align}
	P
	&=
	\mathbb E\!\left\{
	\mathbf i_{\mathrm{tot}}^{\mathsf H}\,
	\Re\{\mathbf Z(\mathbf p)\}\,
	\mathbf i_{\mathrm{tot}}
	\right\} \nonumber\\
	&=
	\mathbb E\!\left\{
	\mathbf s^{\mathsf H}
	\mathbf U^{\mathsf H}\widetilde{\mathbf W}^{\mathsf H}(\mathbf p)\,
	\Re\{\mathbf Z(\mathbf p)\}\,
	\widetilde{\mathbf W}(\mathbf p)\mathbf U
	\mathbf s
	\right\}.
	\label{eq:P_def_mid_mu}
\end{align}
With $\mathbb E\{\mathbf s\mathbf s^{\mathsf H}\}=\mathbf I_K$, the transmit power simplifies to
\begin{align}
	P
	=
	\mathrm{tr}\!\Big(
	\mathbf U^{\mathsf H}\mathbf B(\mathbf p)\mathbf U
	\Big),
	\label{eq:P_diag_mu}
\end{align}
where
\begin{align}
	\mathbf B(\mathbf p)
	=
	\widetilde{\mathbf W}^{\mathsf H}(\mathbf p)\,
	\Re\{\mathbf Z(\mathbf p)\}\,
	\widetilde{\mathbf W}(\mathbf p)
	\label{eq:B_def_mu}
\end{align}
is positive semidefinite and captures the effects of mutual coupling and the movement configuration on the power consumption.

Under the block-diagonal structure in \eqref{eq:Z_blockdiag} and the local mechanical beamforming vectors in \eqref{eq:iCm_def}, $\mathbf B(\mathbf p)$ takes the following diagonal form,
\begin{align}
	\mathbf B(\mathbf p)
	= \mathrm{diag}\big\{b_1(\mathbf p_1),\ldots,b_M(\mathbf p_M)\big\},
	\label{eq:B_diag_mu}
\end{align}
where
\begin{align}
	b_m(\mathbf p_m)
	=
	\widetilde{\mathbf w}_m^{\mathsf H}(\mathbf p_m)\,
	\Re\{\mathbf Z_m(\mathbf p_m)\}\,
	\widetilde{\mathbf w}_m(\mathbf p_m),
	\label{eq:b_m_def_mu}
\end{align}
with
$\widetilde{\mathbf w}_m(\mathbf p_m)=
\big[1,\,-\mathbf w_m^{\mathsf T}(\mathbf p_m)\big]^{\mathsf T}\in\mathbb C^{(N+1)\times 1}$.

Let $P_{\max}$ denote the maximum available transmit power. The transmit power constraint is given by
\begin{align}
	\mathrm{tr}\!\Big(
	\mathbf U^{\mathsf H}\mathbf B(\mathbf p)\mathbf U
	\Big)
	\le P_{\max}.                                           \label{eq:power_constraint_mu}
\end{align}

We aim to jointly optimize the coupler positions and the digital beamforming matrix to maximize the achievable sum rate, subject to the transmit power constraint and the movement constraints. The problem is formulated as
\begin{subequations}\label{prob:P1_mu}
	\begin{align}
		\text{(P1)}~&
		\max_{\{\mathbf p_m\}_{m=1}^{M},\,\mathbf U}\quad
		R(\mathbf p,\mathbf U)                                \label{prob:P1_obj_mu}\\
		\text{s.t.}~~
		& \mathrm{tr}\!\Big(
		\mathbf U^{\mathsf H}\mathbf B(\mathbf p)\mathbf U
		\Big)
		\le P_{\max},                                          \label{prob:P1_power_mu}\\
		& \mathbf p_{n,m}\in\mathcal C_m,
	 \forall n\in\mathcal N, \forall m\in\mathcal M,   \label{prob:P1_region_mu}\\
		& \|\mathbf p_{n,m}-\mathbf p_{n',m}\|_{2}
		\ge d_{\min},
		 \forall n,n'\in\{0\} \cup\mathcal N,\nonumber\\
		 &~~~~~~~~~~~~~~~~~~~~~~~~~~~ n\ne n', m\in\mathcal M,  \label{prob:P1_spacing_mu}
	\end{align}
\end{subequations}
where $\mathcal N=\{1,2,\cdots,N\}$ and $\mathbf p_{0,m}\triangleq \mathbf q_m$ denotes the location of the $m$th active antenna.
Problem (P1) is nonconvex because the sum rate in \eqref{prob:P1_obj_mu} is nonconcave with respect to $(\mathbf p,\mathbf U)$ and the spacing constraints in \eqref{prob:P1_spacing_mu} are nonconvex. In the following, we first derive a closed-form MMSE-based digital precoder for given coupler positions, and then reduce (P1) to a position-only problem that is amenable to distributed optimization.

\subsection{MMSE-Based Digital Beamforming}
\label{subsec:MMSE_Z}
For fixed coupler positions $\mathbf p$, the received signal at user $k$ can be written as
\begin{align}
	y_k = \mathbf g_k^{\mathsf H}(\mathbf p)\mathbf U\mathbf s + n_k,
	\label{45r}
\end{align}
where the effective channel after mechanical beamforming is
\begin{align}
	\mathbf g_k^{\mathsf H}(\mathbf p)
	= \mathbf h_k^{\mathsf T}(\mathbf p)\,\widetilde{\mathbf W}(\mathbf p)
	\in\mathbb C^{1\times M},
\end{align}
and the effective channel matrix is given by
\begin{align}
	\mathbf G(\mathbf p)
	=
	\big[
	\mathbf g_1(\mathbf p),\ldots,\mathbf g_K(\mathbf p)
	\big]^{\mathsf T}
	\in\mathbb C^{K\times M}.
	\label{eq:G_def}
\end{align}

Using the diagonal structure of $\mathbf B(\mathbf p)$ in \eqref{eq:B_diag_mu}, we
introduce the whitened precoder
\begin{align}
	\mathbf F
	= \mathbf B^{1/2}(\mathbf p)\mathbf U
	\in\mathbb C^{M\times K},
	\label{eq:X_def}
\end{align}
with
\begin{align}
	\mathbf B^{1/2}(\mathbf p)
	=
	\mathrm{diag}\big\{\sqrt{b_1(\mathbf p_1)},\ldots,\sqrt{b_M(\mathbf p_M)}\big\}.
\end{align}
Then, we have $\mathbf U=\mathbf B^{-1/2}(\mathbf p)\mathbf F$, and the transmit power constraint \eqref{eq:P_diag_mu} is equivalent to
\begin{align}
	\|\mathbf F\|_F^2 \le P_{\max}.
	\label{eq:X_power}
\end{align}
Substituting $\mathbf U=\mathbf B^{-1/2}(\mathbf p)\mathbf F$ into \eqref{45r} yields
\begin{align}
	y_k
	= \overline{\mathbf g}_k^{\mathsf H}(\mathbf p)\mathbf F\mathbf s + n_k,
\end{align}
where $
\overline{\mathbf g}_k^{\mathsf H}(\mathbf p)
=
\mathbf g_k^{\mathsf H}(\mathbf p)\,\mathbf B^{-1/2}(\mathbf p)$.

For given coupler positions \(\mathbf p\), the FCA system reduces to a conventional downlink multiuser MIMO with  precoder $\mathbf F$ under \eqref{eq:X_power} and effective channel matrix
\begin{align}
	\overline{\mathbf G}(\mathbf p)
	=
	\big[
	\overline{\mathbf g}_1(\mathbf p),\ldots,\overline{\mathbf g}_K(\mathbf p)
	\big]^{\mathsf T}
	= \mathbf G(\mathbf p)\mathbf B^{-1/2}(\mathbf p).
\end{align}
We obtain $\mathbf F$ by solving the regularized least-square
(LS) problem
\begin{align}
	\min_{\mathbf F}\;
	\big\|
	\overline{\mathbf G}(\mathbf p)\mathbf F - \mathbf I_K
	\big\|_F^2
	+ \alpha \|\mathbf F\|_F^2,
	\label{eq:MMSE_LS}
\end{align}
whose solution satisfies
\begin{align}
	\widehat{\mathbf F}_{\rm MMSE}(\mathbf p)
	=
	\overline{\mathbf G}^{\mathsf H}(\mathbf p)
	\Big(
	\overline{\mathbf G}(\mathbf p)\overline{\mathbf G}^{\mathsf H}(\mathbf p)
	+ \alpha \mathbf I_K
	\Big)^{-1}.
	\label{eq:Xhat_MMSE}
\end{align}
where $\alpha = K\sigma^2/P_{\max}$ \cite{nguyen2014mmse}. To enforce \eqref{eq:X_power} with equality, we further
apply a scalar power loading factor
\begin{align}
	\beta(\mathbf p)
	=
	\sqrt{
		\frac{P_{\max}}
		{\big\|\widehat{\mathbf F}_{\rm MMSE}(\mathbf p)\big\|_F^2}
	},
	\label{eq:beta_def}
\end{align}
and set $\mathbf F_{\rm MMSE}(\mathbf p)=\beta(\mathbf p)\widehat{\mathbf F}_{\rm MMSE}(\mathbf p)$. Therefore, the MMSE-based digital precoder is given by
\begin{align}
	\mathbf U_{\rm MMSE}(\mathbf p)
	= \mathbf B^{-1/2}(\mathbf p)\,\mathbf F_{\rm MMSE}(\mathbf p)
	\in\mathbb C^{M\times K}.
	\label{eq:Z_MMSE}
\end{align}
This construction guarantees
\begin{align}
	\mathrm{tr}\big(\mathbf U_{\rm MMSE}^{\mathsf H}\mathbf B(\mathbf p)\mathbf U_{\rm MMSE}\big)
	=
	\|\mathbf F_{\rm MMSE}(\mathbf p)\|_F^2
	= P_{\max}.
\end{align}

\subsection{Coupler Position Optimization Algorithm} 
Since the MMSE-based digital precoder $\mathbf U_{\rm MMSE}(\mathbf p)$ is
available in closed form for any given $\mathbf p$ and already satisfies the
power constraint in \eqref{prob:P1_power_mu}, we eliminate $\mathbf U$ from
problem (P1). Specifically, by substituting $\mathbf U=\mathbf U_{\rm MMSE}(\mathbf p)$ into
\eqref{eq:SINR_mu}, the SINR of user $k$ becomes
\begin{align}
	\gamma_k^{\rm MMSE}(\mathbf p)
	=
	\frac{\big|
		\mathbf h_k^{\mathsf T}(\mathbf p)\,
		\widetilde{\mathbf W}(\mathbf p)\,
		\mathbf u_{k}^{\rm MMSE}(\mathbf p)
		\big|^2}
	{\displaystyle
		\sum_{j\in\mathcal K,\,j\neq k}
		\big|
		\mathbf h_k^{\mathsf T}(\mathbf p)\,
		\widetilde{\mathbf W}(\mathbf p)\,
		\mathbf u_{j}^{\rm MMSE}(\mathbf p)
		\big|^2
		+ \sigma_k^2},
\end{align}
where $\mathbf u_{k}^{\rm MMSE}(\mathbf p)$ is the $k$-th column of
$\mathbf U_{\rm MMSE}(\mathbf p)$. The corresponding achievable sum rate is
\begin{align}
	R_{\rm MMSE}(\mathbf p)
	= \sum_{k=1}^{K}
	\log_2\big(1+\gamma_k^{\rm MMSE}(\mathbf p)\big).    \label{eq:R_MMSE_p}
\end{align}
This leads to the following position-only optimization problem,
\begin{subequations}\label{prob:P2}
	\begin{align}
		\text{(P2)}~&
		\max_{\{\mathbf p_m\}_{m=1}^{M}}\quad
		R_{\rm MMSE}(\mathbf p)                           \label{prob:P2_obj}\\
		\text{s.t.}~&
		\eqref{prob:P1_region_mu},\eqref{prob:P1_spacing_mu}.
	\end{align}
\end{subequations}
Problem (P2) is still nonconvex due to the complicated dependence of
$R_{\rm MMSE}(\mathbf p)$ on $\mathbf p$ and the nonconvex spacing constraints in
\eqref{prob:P1_spacing_mu}. However, the two constraints
\eqref{prob:P1_region_mu} and \eqref{prob:P1_spacing_mu} are separable across
different FCAs, which enables a distributed optimization algorithm.

Let
\begin{align}
	\mathcal S_m
	&=
	\Big\{
	\mathbf p_m \,\big|\,
	\mathbf p_{n,m}\in\mathcal C_m,\;
	\|\mathbf p_{n,m}-\mathbf p_{n',m}\|_2 \ge d_{\min},\nonumber\\
	&\hspace{5mm}\forall n,n'\in\{0\} \cup\mathcal N,~n\neq n'
	\Big\},
\end{align}
denote the local feasible set of coupler positions at the $m$-th FCA.
Due to the spacing constraints, $\mathcal S_m$ is in general nonconvex. To apply
the SCA framework with convergence guarantees, we construct at each iteration a
convex inner approximation of $\mathcal S_m$.
Specifically, for each pair $(n,n')$ with $n\neq n'$, we define
\begin{align}
	d_{n,n',m}(\mathbf p_m)
	=
	\big\|\mathbf p_{n,m}-\mathbf p_{n',m}\big\|_2^2,
\end{align}
so that the spacing constraint is equivalent to
$-d_{n,n',m}(\mathbf p_m)\le -d_{\min}^2$.
Since $-d_{n,n',m}(\mathbf p_m)$ is concave, its first-order Taylor
approximation at $\mathbf p_m^{(t)}$ yields a global affine upper bound. Denote
\begin{align}
	&\widetilde d_{n,n',m}(\mathbf p_m;\mathbf p_m^{(t)})\nonumber\\
	&=
	-d_{n,n',m}(\mathbf p_m^{(t)})
	-\nabla_{\mathbf p_m} d_{n,n',m}(\mathbf p_m^{(t)})^{\mathsf T}
	\big(\mathbf p_m-\mathbf p_m^{(t)}\big),
\end{align}
where the gradient can be obtained in closed form. At iteration $t$, we
approximate the spacing constraint by
\begin{align}
	\widetilde d_{n,n',m}(\mathbf p_m;\mathbf p_m^{(t)})
	\le -d_{\min}^2,
	\forall n\neq n',~ n,n'\in\{0\}\cup\mathcal N.
\end{align}
The resulting convex inner approximation of $\mathcal S_m$ is
\begin{align}
	\mathcal S_m^{(t)}
	&=
	\Big\{
	\mathbf p_m \,\big|\,
	\mathbf p_{n,m}\in\mathcal C_m,\;
	\widetilde d_{n,n',m}(\mathbf p_m;\mathbf p_m^{(t)})\le -d_{\min}^2,
	\nonumber\\
	&\hspace{10mm}\forall n,n'\in\{0\} \cup\mathcal N,~n\neq n'
	\Big\},
\end{align}
which satisfies $\mathcal S_m^{(t)}\subseteq\mathcal S_m$ for all $t$.

For problem (P2), the feasible set is separable across the $M$ FCAs
through $\{\mathcal S_m\}_{m=1}^M$, while the objective function
$R_{\rm MMSE}(\mathbf p)$ couples all antennas through the effective channels
and the MMSE-based digital precoder. Let
\begin{align}
	J(\mathbf p) = R_{\rm MMSE}(\mathbf p)
\end{align}
denote the sum-rate objective in (P2). Given $\mathbf p^{(t)}$ at iteration
$t$, we consider a first-order approximation of $J(\mathbf p)$ with respect to
$\mathbf p_m$ around $\mathbf p_m^{(t)}$, and add a quadratic regularization term.
For each $m\in\mathcal M$, we define the local surrogate
function as
\begin{align}
	\widetilde J_m(\mathbf p_m;\mathbf p^{(t)})
	&=
	J(\mathbf p^{(t)})
	+ \big(\nabla_{\mathbf p_m} J(\mathbf p^{(t)})\big)^{\mathsf T}
	\big(\mathbf p_m - \mathbf p_m^{(t)}\big) \nonumber\\
	&\quad
	- \frac{\eta_m}{2}\big\|\mathbf p_m - \mathbf p_m^{(t)}\big\|_2^2,
	\label{eq:local_surrogate}
\end{align}
where $\eta_m>0$ is a design parameter no smaller than the Lipschitz constant of
$\nabla_{\mathbf p_m} J(\mathbf p)$ to ensure monotonic improvement.

The gradient of the sum rate with respect to $\mathbf p_m$ is denoted by
\begin{align}\label{69}
	\mathbf g_m(\mathbf p)
	= \nabla_{\mathbf p_m} J(\mathbf p)\in\mathbb R^{2N},
\end{align}
which can be obtained by applying the chain rule to \eqref{eq:R_MMSE_p}. 

For fixed $\mathbf p^{(t)}$ and each $m$, we consider the following local
strongly concave maximization problem
\begin{subequations}\label{prob:P3_m}
	\begin{align}
		\text{(P3-$m$)}~\max_{\mathbf p_m}\quad
		& \widetilde J_m(\mathbf p_m;\mathbf p^{(t)}) \label{prob:P3_m_obj}\\
		\text{s.t.}\quad
		& \mathbf p_m\in\mathcal S_m^{(t)}. \label{prob:P3_m_feas}
	\end{align}
\end{subequations}
By substituting \eqref{69} into \eqref{eq:local_surrogate} and dropping the
constant term $J(\mathbf p^{(t)})$, (P3-$m$) is equivalent to
\begin{align} \label{unc3}
	\max_{\mathbf p_m\in\mathcal S_m^{(t)}}\;
	\mathbf g_m^{\mathsf T}(\mathbf p^{(t)})
	\big(\mathbf p_m - \mathbf p_m^{(t)}\big)
	- \frac{\eta_m}{2}
	\big\|\mathbf p_m - \mathbf p_m^{(t)}\big\|_2^2.
\end{align}
Ignoring the constraint in \eqref{unc3} yields the following unconstrained maximizer
\begin{align}
	\mathbf p_m^{\rm uncon}
	= \mathbf p_m^{(t)} + \frac{1}{\eta_m}\mathbf g_m(\mathbf p^{(t)}).
\end{align}
Then, the solution of (P3-$m$) is obtained by projecting $\mathbf p_m^{\rm uncon}$
onto $\mathcal S_m^{(t)}$,
\begin{align}
	\widehat{\mathbf p}_m^{(t+1)}
	&= \mathcal P_{\mathcal S_m^{(t)}}\Big(
	\mathbf p_m^{(t)} + \tfrac{1}{\eta_m}\mathbf g_m(\mathbf p^{(t)})
	\Big),
	\label{eq:local_proj_MMSE}
\end{align}
where $\mathcal P_{\mathcal S_m^{(t)}}(\cdot)$ denotes the Euclidean projection onto
$\mathcal S_m^{(t)}$. This projection can be performed locally by enforcing
\eqref{prob:P1_region_mu} and the linearized spacing constraints.

To further improve robustness, we adopt the relaxation step
\begin{align}
	\mathbf p_m^{(t+1)}
	= \mathbf p_m^{(t)}
	+ \alpha_m^{(t)}\big(
	\widehat{\mathbf p}_m^{(t+1)} - \mathbf p_m^{(t)}\big),
	\label{eq:relax_update}
\end{align}
where $\alpha_m^{(t)}\in(0,1]$ satisfies standard diminishing conditions, i.e.,
$\sum_{t} \alpha_m^{(t)} = \infty$ and
$\sum_{t} (\alpha_m^{(t)})^2 < \infty$ \cite{4749425}.

\begin{algorithm}[t]
	\caption{Distributed Coupler Position Optimization Algorithm}
	\label{alg:CP_MMSE}
	\begin{algorithmic}[1]
		\STATE \textbf{Input:} $P_{\max}$, $\{\mathcal C_m\}_{m=1}^{M}$, $d_{\min}$,
		$\{\eta_m\}_{m=1}^{M}$, $\{\alpha_m^{(t)}\}$, tolerance $\varepsilon>0$.
		\STATE \textbf{Initialization:} Choose a feasible $\mathbf p^{(0)}$
		satisfying \eqref{prob:P1_region_mu}--\eqref{prob:P1_spacing_mu}; set $t=0$.
		\STATE CPU computes $\overline{\mathbf G}(\mathbf p^{(0)})$ and
		$\mathbf U_{\rm MMSE}(\mathbf p^{(0)})$
		according to \eqref{eq:Xhat_MMSE}--\eqref{eq:Z_MMSE}, and evaluates
		$R^{(0)} = R_{\rm MMSE}(\mathbf p^{(0)})$.
		\REPEAT
		\STATE CPU computes $\mathbf g_m(\mathbf p^{(t)})$ for all $m\in\mathcal M$
		and broadcasts $\{\mathbf g_m(\mathbf p^{(t)})\}_{m=1}^M$.
		\FOR{$m=1$ to $M$ \textbf{(in parallel)}}
		\STATE $\mathbf p_m^{\rm uncon}
		= \mathbf p_m^{(t)} + \tfrac{1}{\eta_m}\mathbf g_m(\mathbf p^{(t)})$.
		\STATE $\widehat{\mathbf p}_m^{(t+1)}
		=\mathcal P_{\mathcal S_m^{(t)}}(\mathbf p_m^{\rm uncon})$
		via \eqref{eq:local_proj_MMSE}.
		\STATE Update $\mathbf p_m^{(t+1)}$ via \eqref{eq:relax_update}.
		\ENDFOR
		\STATE CPU forms $\mathbf p^{(t+1)}$ by stacking
		$\{\mathbf p_m^{(t+1)}\}_{m=1}^{M}$.
		\STATE CPU recomputes $\overline{\mathbf G}(\mathbf p^{(t+1)})$ and
		updates $\mathbf U_{\rm MMSE}(\mathbf p^{(t+1)})$
		according to \eqref{eq:Xhat_MMSE}--\eqref{eq:Z_MMSE}.
		\STATE CPU evaluates $R^{(t+1)} = R_{\rm MMSE}(\mathbf p^{(t+1)})$
		via \eqref{eq:R_MMSE_p}.
		\STATE $t \leftarrow t+1$.
		\UNTIL{$\big|R^{(t)}-R^{(t-1)}\big|/R^{(t-1)} \le \varepsilon$}
		\STATE \textbf{Output:} $\mathbf p^{\star} = \mathbf p^{(t)}$ and
		$\mathbf U_{\rm MMSE}^{\star}= \mathbf U_{\rm MMSE}(\mathbf p^{\star})$.
	\end{algorithmic}
\end{algorithm}

Algorithm~\ref{alg:CP_MMSE} summarizes the distributed coupler position optimization procedure based on a distributed SCA scheme, where each FCA maximizes a strongly concave local surrogate of the global sum rate. Under standard smoothness assumptions on $J(\mathbf p)$ and a diminishing step size sequence, the generated sequence $\{\mathbf p^{(t)}\}$ yields nondecreasing sum rate $J(\mathbf p^{(t)})$, and every limit point of $\{\mathbf p^{(t)}\}$ is a stationary solution of problem (P2). The per-iteration computational complexity is dominated by the computation of the MMSE-based precoder and the gradients, and is given by
\(\mathcal O(K^2 M + K^3 + K M N + M N^2)\).
Let \(T_{\max}\) denote the number of iterations required by the stopping criterion in Algorithm~\ref{alg:CP_MMSE}. The overall complexity therefore scales as
\(\mathcal O\!\big(T_{\max}(K^2 M + K^3 + K M N + M N^2)\big)\).
At each iteration, the CPU broadcasts the local gradients to all LPUs, and each LPU uploads its updated coupler-position vector to the CPU. Hence, the communication overhead scales on the order of \(MN\) scalar parameters.

\section{Channel Estimation for Flexible Coupler Systems}
\begin{figure}[t!]
	\centering
	\setlength{\abovecaptionskip}{0.cm}
	\includegraphics[width=3.16in]{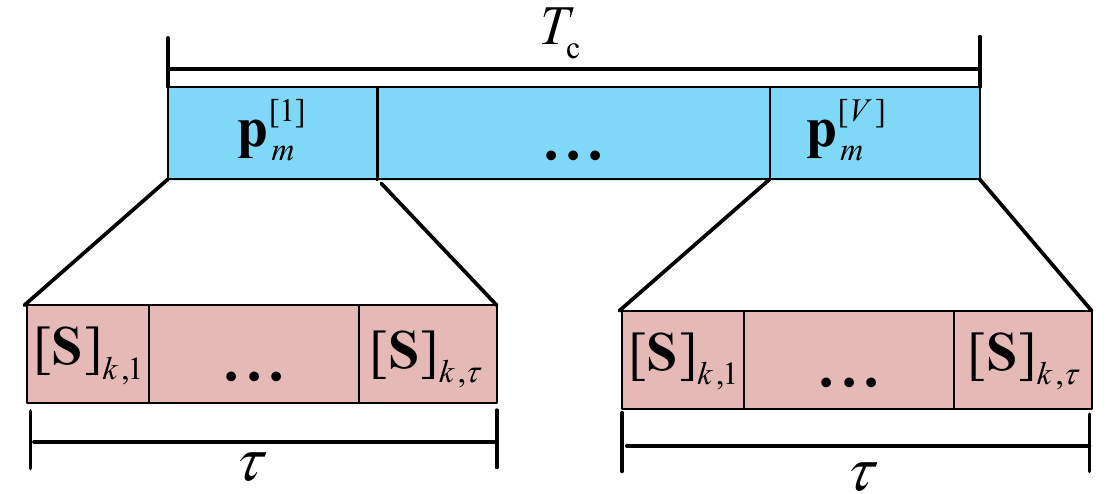}
	\caption{Structured pilot pattern in the time domain.}
	\label{protocol}
		\vspace{-0.3cm}
\end{figure}

As shown in Fig.~\ref{protocol}, we propose a channel estimation protocol for FCA systems. Specifically, the channel training length $T_c$ (in time slots) is divided into $V$ pilot blocks indexed by $v\in\{1,\ldots,V\}$. 
At pilot block $v$, FCA $m$ sets the positions of its associated $N$ couplers to $\mathbf p_m^{[v]}$.
We assume that all user channels remain constant over $T_c = V\tau$, where $\tau$ denotes the number of time slots in each pilot block. Therefore, the coupler position vector $\mathbf p_m^{[v]}$ remains unchanged during the $\tau$ time slots of the $v$-th block, and it varies across different blocks.
Each user transmits a known pilot sequence $[\mathbf S]_{k,:}\in\mathbb C^{1\times \tau}$ to the BS, and the same pilot is repeated over the $V$ blocks. Let pilot matrix $\mathbf S\in\mathbb C^{K\times \tau}$ satisfy $\mathbf S\mathbf S^{\mathsf H}=\tau \mathbf I_K$.

Based on the proposed protocol, we first develop a pilot-assisted centralized channel estimation algorithm with high estimation accuracy and then present a pilot-assisted distributed channel estimation algorithm for FCA systems. 

\subsection{Centralized Channel Estimation}
First, we introduce the centralized channel estimation algorithm. Let 
\begin{align}
	\mathbf h_{m,k}(\mathbf p_m^{[v]})
	=
	\begin{bmatrix}
		[\mathbf h_{\mathrm A,k}]_{m}\\
		\mathbf h_{\mathrm C,k}(\mathbf p_m^{[v]})
	\end{bmatrix}
	\in\mathbb C^{(N+1)\times 1},
	\label{eq:CE_h_local_def}
\end{align}
denote the local channel from user $k$ to the $m$-th FCA under coupler position $\mathbf p_m^{[v]}$, where $\mathbf h_{\mathrm C,k}(\mathbf p_m^{[v]})
=
\sum_{\ell=1}^{L_k} \alpha_{k,\ell}
\mathbf a_{\mathrm C,k,\ell}(\mathbf p_m^{[v]})\in\mathbb C^{N\times 1}$ collects the $N$ coupler to user-$k$ channels
associated with $\mathbf p_m^{[v]}$.
Under the proposed pilot pattern, the received signal at FCA $m$ is given by
\begin{align}
	\mathbf y_m^{[v]}
	=
	\widetilde{\mathbf w}_m^{\mathsf T}(\mathbf p_m^{[v]})
	\mathbf H_m(\mathbf p_m^{[v]})
	\mathbf S
	+\mathbf n_m^{[v]}\in\mathbb C^{1\times \tau},
	\label{eq:CEc_local_pilot}
\end{align}
where $\mathbf H_m(\mathbf p_m^{[v]})=[\mathbf h_{m,1}(\mathbf p_m^{[v]}),\ldots,\mathbf h_{m,K}(\mathbf p_m^{[v]})]\in\mathbb C^{(N+1)\times K}$ denotes the channel matrix between the $m$-th FCA and the users, and
$\mathbf n_m^{[v]}\in\mathbb C^{1\times \tau}$ is the received Gaussian noise
vector.

In the centralized channel estimation scheme, for each pilot block $v$, every FCA sends its received signal $\mathbf y_m^{[v]}$ to the CPU.
The CPU stacks the received pilots across the $M$ antennas as
\begin{align}
	\mathbf Y^{[v]}
	=
	\big[
	(\mathbf y_1^{[v]})^{\mathsf T},\ldots,(\mathbf y_M^{[v]})^{\mathsf T}
	\big]^{\mathsf T}
	\in\mathbb C^{M\times \tau}.
	\label{eq:CEc_Yv_def}
\end{align}

Based on \eqref{eq:CEc_Yv_def}, the CPU performs pilot correlation as
\begin{align}
	\widehat{\mathbf G}^{[v]}
	=
	\frac{1}{\tau}\mathbf Y^{[v]}\mathbf S^{\mathsf H}.
	\label{eq:CEc_Ghat}
\end{align}
Consequently, for each user $k$, we obtain the stacked effective channel observation vector as 
\begin{align}
	\widehat{\mathbf y}_k
	=
	\big[
	(\widehat{\mathbf g}_k^{[1]})^{\mathsf T},\ldots,(\widehat{\mathbf g}_k^{[V]})^{\mathsf T}
	\big]^{\mathsf T}
	\in\mathbb C^{M V\times 1},
	\label{eq:CEc_yk}
\end{align}
where $\widehat{\mathbf g}_k^{[v]}
= [\widehat{\mathbf G}^{[v]}]_{:,k}
\in\mathbb C^{M\times 1}$.

Moreover, according to \eqref{vty_mu}, the effective channel from user $k$ to FCA $m$ at signal block $v$ can be written as
\begin{subequations}
	\begin{align}
		g_{m,k}(\mathbf p_m^{[v]})&	=
		\widetilde{\mathbf w}_m^{\mathsf T}(\mathbf p_m^{[v]})
		\mathbf h_{m,k}(\mathbf p_m^{[v]})\\
		&=
		\sum_{\ell=1}^{L_k}\alpha_{k,\ell}\,b_m(\phi_{k,\ell};\mathbf p_m^{[v]}),
		\label{eq:CEc_geom}
	\end{align}
\end{subequations}
where
\begin{align}
	b_m(\phi_{k,\ell};\mathbf p_m^{[v]})
	=
	[\mathbf a_y(\phi_{k,\ell})]_{m}
	-
	\mathbf w_m^{\mathsf T}(\mathbf p_m^{[v]})\,\mathbf a_{\mathrm C,k,\ell}(\phi_{k,\ell};\mathbf p_m^{[v]}).
	\label{eq:CEc_bm}
\end{align}

To efficiently estimate $\{\alpha_{k,\ell}\}$ via compressed sensing, we approximate the azimuth domain by a uniform grid $\mathcal G=\{\phi_1,\ldots,\phi_G\}$, where $G\gg L_k$ denotes the number of grid points.
For candidate angle $\phi$ and pilot block $v$, we define
\begin{align}
	\mathbf b(\phi;\mathbf p^{[v]})
	=
	\big[
	b_1(\phi;\mathbf p_1^{[v]}),\ldots,b_M(\phi;\mathbf p_M^{[v]})
	\big]^{\mathsf T}
	\in\mathbb C^{M\times 1},
	\label{eq:CEc_bvec}
\end{align}
where $\mathbf p^{[v]}=\{\mathbf p_m^{[v]}\}_{m=1}^{M}$.
Then, the dictionary is constructed as
\begin{align}
	\mathbf A
	=
	\big[
	\mathbf a(\phi_1),\ldots,\mathbf a(\phi_G)
	\big]
	\in\mathbb C^{M V\times G},
	\label{eq:CEc_dict}
\end{align}
where
\begin{align}
	\mathbf a(\phi)
	=
	\big[
	\mathbf b(\phi;\mathbf p^{[1]})^{\mathsf T},\ldots,\mathbf b(\phi;\mathbf p^{[V]})^{\mathsf T}
	\big]^{\mathsf T}
	\in\mathbb C^{M V\times 1}.
\end{align}

Since $\mathbf S\mathbf S^{\mathsf H}=\tau \mathbf I_K$, for each user $k$, the stacked observation in \eqref{eq:CEc_yk} can be expressed as the following sparse signal model
\begin{align}
	\widehat{\mathbf y}_k=\mathbf A\mathbf x_k+\widetilde{\mathbf n}_k,
	\label{eq:CEc_sparse}
\end{align}
where $\widetilde{\mathbf n}_k$ denotes the received Gaussian noise vector, $\mathbf x_k\in\mathbb C^{G\times 1}$ is $L_k$ sparse with $L_k$ nonzero elements that correspond to the $L_k$ propagation paths between user $k$ and the BS. In particular, the support of $\mathbf x_k$ specifies the $L_k$ grid indices that approximate azimuth angles $\{\phi_{k,\ell}\}_{\ell=1}^{L_k}$, and the associated nonzero coefficients represent the corresponding path gains.

Using orthogonal matching pursuit (OMP) \cite{7476849}, we perform exactly $L_k$ selections and obtain the support set, denoted by $\widetilde{\Omega}_k=\{\widehat{\jmath}_{k,1},\ldots,\widehat{\jmath}_{k,L_k}\}$. The estimated angles are obtained by mapping the selected indices to the predefined grid as
\begin{align}
	\widehat{\phi}_{k,\ell}= \phi_{\widehat{\jmath}_{k,\ell}}, \ell=1,\ldots,L_k.
	\label{eq:CEc_phi_hat}
\end{align}

After obtaining ${\widehat{\phi}_{k,\ell}}$, we estimate the path gains using the LS technique, the resulting solution is given by
\begin{align}
	\widehat{\boldsymbol\alpha}_k
	=
	(\mathbf A_k^{\mathsf H}\mathbf A_k)^{-1}\mathbf A_k^{\mathsf H}\widehat{\mathbf y}_k
	\in\mathbb C^{L_k\times 1},
	\label{eq:CEc_alpha_hat}
\end{align}
where
\begin{align}
	\mathbf A_k
	=
	\big[
	\mathbf a(\widehat{\phi}_{k,1}),\ldots,\mathbf a(\widehat{\phi}_{k,L_k})
	\big]
	\in\mathbb C^{M V\times L_k}.
\end{align}

After obtaining $\{(\widehat{\phi}_{k,\ell},\widehat{\alpha}_{k,\ell})\}$, the effective channel under any coupler position $\mathbf p_m$ can be reconstructed as
\begin{align}
	\widehat g_{m,k}(\mathbf p_m)
=
\sum_{\ell=1}^{L_k}
[\widehat{\boldsymbol\alpha}_k]_{\ell}\,
b_m(\widehat\phi_{k,\ell};\mathbf p_m).
	\label{eq:CEc_recon}
\end{align}

The main steps of the proposed centralized channel estimation procedure are summarized in Algorithm~\ref{alg:CE_centralized}.
The computational complexity of Algorithm~\ref{alg:CE_centralized} is dominated by the OMP-based sparse recovery and scales as $\mathcal{O}\!\big(\sum_{k=1}^{K} L_k\, M V\, G\big)$, while constructing the dictionary incurs an additional $\mathcal{O}(M V G N)$ operations. Therefore, the overall complexity is $\mathcal{O}\!\big(\sum_{k=1}^{K} L_k\, M V\, G + M V G N\big)$. In terms of communication overhead, Algorithm 2 requires each FCA to upload its length-$\tau$ received pilot vectors $\{\mathbf y_m^{[v]}\}_{v=1}^{V}$ to the CPU over $V$ pilot blocks, resulting in a communication overhead on the order of $\mathcal O(MV\tau)$.
The centralized channel estimation scheme provides high estimation accuracy since the CPU jointly processes the pilot-correlated observations stacked over all FCAs and all pilot blocks.

\begin{algorithm}[t]
	\caption{Centralized channel estimation for FCA system}
	\label{alg:CE_centralized}
	\begin{algorithmic}[1]
		\STATE \textbf{Input} $\mathbf S$, $\tau$,  $\mathcal G$,  $\{L_k\}_{k=1}^{K}$,
	    $\{\mathbf p_m^{[v]}\}_{m=1,v=1}^{M,V}$.
		\FOR{$v=1$ to $V$}
		\STATE Each FCA $m$ applies $\mathbf p_m^{[v]}$ and receives $\mathbf y_m^{[v]}$ in \eqref{eq:CEc_local_pilot}.
		\STATE Each FCA $m$ uploads $\mathbf y_m^{[v]}$ to the CPU.
		\STATE The CPU forms $\mathbf Y^{[v]}$ in \eqref{eq:CEc_Yv_def} and computes $\widehat{\mathbf G}^{[v]}$ in \eqref{eq:CEc_Ghat}.
		\ENDFOR
		\STATE The CPU builds the dictionary $\mathbf A$ in \eqref{eq:CEc_dict} using \eqref{eq:CEc_bvec}.
		\FOR{each user $k$ at the CPU}
		\STATE Form $\widehat{\mathbf y}_k$ in \eqref{eq:CEc_yk}.
		\STATE Run OMP on \eqref{eq:CEc_sparse} with $L_k$ selections and obtain $\widetilde{\Omega}_k$.
		\STATE Obtain $\{\widehat{\phi}_{k,\ell}\}_{\ell=1}^{L_k}$ by \eqref{eq:CEc_phi_hat}.
		\STATE Compute $\widehat{\boldsymbol\alpha}_k$ in \eqref{eq:CEc_alpha_hat}.
		\ENDFOR
		\STATE \textbf{Output} $\{\widehat{\phi}_{k,\ell},\widehat{\alpha}_{k,\ell}\}$ and $\widehat g_{m,k}(\mathbf p_m)$.
	\end{algorithmic}
\end{algorithm}

\subsection{Distributed Channel Estimation}
The centralized channel estimation approach suffers from prohibitively high computational complexity, processing cost, and latency at the CPU, especially when the BS is equipped with an extremely large FCA array with a large number of active antennas $M$ and a vast number of coupler candidate positions. We aim to design a distributed channel estimation scheme for the FCA array that achieves accuracy close to the centralized baseline, while reducing the CPU computational burden.
\subsubsection{Processing at local nodes}
By applying pilot correlation to \eqref{eq:CEc_local_pilot}, we obtain
\begin{align}
	\widehat g_{m,k}^{[v]}
	=
	\frac{1}{\tau}\mathbf y_m^{[v]}\mathbf S^{\mathsf H}\mathbf e_k,
	\label{eq:CE_LS_gmk}
\end{align}
where $\mathbf e_k$ is the $k$-th standard basis vector. Stacking $\{\widehat g_{m,k}^{[v]}\}_{v=1}^{V}$ yields the local observation vector for user $k$ as
\begin{align}
	\widehat{\mathbf y}_{m,k}
	=
	\big[\widehat g_{m,k}^{[1]},\ldots,\widehat g_{m,k}^{[V]}\big]^{\mathsf T}
	\in\mathbb C^{V\times 1}.
	\label{eq:CE_stack_local6}
\end{align}

Similar to the centralized channel estimation step, we discretize the azimuth domain into $G\gg L_k$ grid points $\{\phi_1,\ldots,\phi_G\}$ and define the local dictionary at FCA $m$ as
\begin{align}
	\mathbf A_m
	=
	\big[
	\mathbf a_m(\phi_1),\ldots,\mathbf a_m(\phi_G)
	\big]
	\in\mathbb C^{V\times G},
	\label{eq:CE_dict_local}
\end{align}
where $[\mathbf a_m(\phi)]_v=b_m(\phi;\mathbf p_m^{[v]})$ for $v=1,\ldots,V$.
Thus, \eqref{eq:CE_stack_local6} can be expressed as the sparse model
\begin{align}
	\widehat{\mathbf y}_{m,k}
	=
	\mathbf A_m\,\mathbf x_{m,k}
	+\mathbf n_{m,k},
	\label{eq:CE_sparse_model}
\end{align}
where $\mathbf x_{m,k}\in\mathbb C^{G\times 1}$ is sparse and its support encodes the azimuth angles of user $k$ on the grid. The support is common across $m$ since these angles are global parameters shared by all FCAs. Moreover, $\mathbf n_{m,k}\!\sim\!\mathcal{CN}(\mathbf 0,\sigma_{m,\mathrm{eff}}^2\mathbf I_V)$ denotes the effective noise vector after pilot correlation and stacking over the $V$ pilot blocks, with $\sigma_{m,\mathrm{eff}}^2=\sigma_m^2/\tau$.

To convey support information with low overhead, FCA $m$ computes a matched-filter proxy $\mathbf c_{m,k}=\mathbf A_m^{\mathsf H}\widehat{\mathbf y}_{m,k}$ and its normalized energy
$\rho_{m,k}(j)=\frac{|[\mathbf c_{m,k}]_j|^2}{\|\mathbf a_m(\phi_j)\|_2^2+\epsilon_{\mathrm n}}$, $j=1,\ldots,G$,
where $\epsilon_{\mathrm n}>0$ avoids numerical issues when $\|\mathbf a_m(\phi_j)\|_2^2$ is small.
It then selects a candidate support set via noise-calibrated hard thresholding \cite{duarte2005distributed}:
\begin{align}
	\mathcal I_{m,k}
	=
	\Big\{
	j\in\{1,\ldots,G\}\ \big|\ 
	\rho_{m,k}(j)\ge \eta\,\sigma_{m,\mathrm{eff}}^2
	\Big\},
	\label{eq:CE_hard_set}
\end{align}
where $\eta>0$ controls the pruning aggressiveness. Finally, FCA $m$ uploads only the index-value pairs $\mathcal U_{m,k}=\{(j,\rho_{m,k}(j))\}_{j\in\mathcal I_{m,k}}$ to the CPU. Since the azimuth domain is sparse, the uploaded list is typically low-dimensional, i.e., $|\mathcal I_{m,k}|\ll G$ in typical multipath channels.

\subsubsection{Processing at central node}
Since azimuth angles $\{\phi_{k,\ell}\}_{\ell=1}^{L_k}$ are path-level geometric parameters and thus common to all FCAs, they are global parameters shared across different FCAs. Local index set $\{\mathcal I_{m,k}\}$ provides complementary evidence of this common support on the grid. Accordingly, the CPU forms an aggregated grid score for each $j\in\{1,\ldots,G\}$ as
\begin{align}
	s_k(j)
	=
	\sum_{m=1}^{M}
	\mathbf 1(j\in\mathcal I_{m,k})\,\rho_{m,k}(j), 
 j=1,\ldots,G,
	\label{eq:CE_score}
\end{align}
where $\mathbf 1(j\in\mathcal I_{m,k})$ equals $1$ if $j\in\mathcal I_{m,k}$ and equals $0$ otherwise, so that only the locally retained indices contribute to $s_k(j)$.

Then, the CPU estimates the support by selecting the $L_k$ grid points that yield the largest aggregated score, namely,
\begin{align}
	\widehat{\Omega}_k
	=
	\arg\max_{\Omega\subseteq\{1,\ldots,G\},\,|\Omega|=L_k}
	\ \sum_{j\in\Omega} s_k(j).
	\label{eq:CE_support_knownL}
\end{align}
Here, $\widehat{\Omega}_k=\{\widehat{\jmath}_{k,1},\ldots,\widehat{\jmath}_{k,L_k}\}$ denotes the selected indices after sorting in descending order such that
$s_k(\widehat{\jmath}_{k,1})\ge \cdots \ge s_k(\widehat{\jmath}_{k,L_k})$.
The estimated azimuth angles are then obtained by mapping the selected grid indices to the corresponding grid points as
\begin{align}
	\widehat{\phi}_{k,\ell}
	=\phi_{\widehat{\jmath}_{k,\ell}},
 \ell=1,\ldots,L_k.
	\label{eq:CE_phi_hat}
\end{align}

\subsubsection{Post-processing at local nodes}
The CPU broadcasts the estimated support set, $\widehat{\Omega}_k$, to all FCAs.
Given $\widehat{\Omega}_k$, FCA $m$ restricts its local dictionary to the columns indexed by $\widehat{\Omega}_k$, namely,
\begin{align}
	\mathbf A_{m,k}^{\mathrm g}
	=
	\big[
	\mathbf a_m(\phi_j)
	\big]_{j\in\widehat{\Omega}_k}
	\in\mathbb C^{V\times L_k},
	\label{eq:CE_subdict_global}
\end{align}
where $\mathbf a_m(\phi_j)\in\mathbb C^{V\times 1}$ is the $m$-th local steering vector across the $V$ pilot blocks associated with grid angle $\phi_j$.
Then, the local sparse recovery problem reduces to estimating the complex path gains on the reduced support.
Specifically, local stacked observation $\widehat{\mathbf y}_{m,k}\in\mathbb C^{V\times 1}$ follows the following low-dimensional linear model,
\begin{align}
	\widehat{\mathbf y}_{m,k}
	=
	\mathbf A_{m,k}^{\mathrm g}\,\boldsymbol\alpha_k
	+\mathbf n_{m,k},
	\label{eq:CE_model_support}
\end{align}
where  $\boldsymbol\alpha_k\in\mathbb C^{L_k\times 1}$ collects the path gains aligned with the ordering of indices in $\widehat{\Omega}_k$.
The reduced model allows each FCA to contribute to estimating the path gains using only a few coefficients, instead of processing the full $G$-dimensional sparse vector.
\begin{algorithm}[t!]
	\caption{Distributed channel estimation for FCA systems}
	\label{alg:dist_ce_fc_proxy}
	\begin{algorithmic}[1]
		\STATE \textbf{Input} $\mathbf S$, $\tau$, $V$, grid $\mathcal G$,
		$\mathbf p_m^{[v]}$,
		$\eta$, $\epsilon_{\mathrm n}$, and $\epsilon_k$.
		\STATE \textbf{Processing at local nodes}
		\FOR{$v=1$ to $V$}
		\STATE For each FCA $m$ \textbf{in parallel}, apply $\mathbf p_m^{[v]}$ and receive $\mathbf y_m^{[v]}$.
		\STATE For each FCA $m$ \textbf{in parallel}, compute $\widehat g_{m,k}^{[v]}$ for all $k$ via \eqref{eq:CE_LS_gmk}.
		\ENDFOR
		\FOR{each FCA $m$ \textbf{in parallel} and each user $k$}
		\STATE Form $\widehat{\mathbf y}_{m,k}$ via \eqref{eq:CE_stack_local6} and build $\mathbf A_m$ via \eqref{eq:CE_dict_local}.
		\STATE Compute $\mathbf c_{m,k}$ and $\rho_{m,k}(j)$.
		\STATE Compute $\mathcal I_{m,k}$ via \eqref{eq:CE_hard_set} and upload $\mathcal U_{m,k}$  to the CPU.
		\ENDFOR
		\STATE \textbf{Processing at central node}
		\FOR{each user $k$ at the CPU}
		\STATE Compute $s_k(j)$ via \eqref{eq:CE_score} and obtain $\widehat{\Omega}_k$ via \eqref{eq:CE_support_knownL}.
		\STATE Obtain $\{\widehat{\phi}_{k,\ell}\}$ via \eqref{eq:CE_phi_hat} and broadcast $\widehat{\Omega}_k$.
		\ENDFOR
		\STATE \textbf{Post-processing at local nodes}
		\FOR{each FCA $m$ \textbf{in parallel} and each user $k$}
		\STATE Form $\mathbf A_{m,k}^{\mathrm g}$ via \eqref{eq:CE_subdict_global}.
		\STATE Compute $(\mathbf R_{m,k},\mathbf q_{m,k})$ via \eqref{eq:CE_suff_stats} and upload them to the CPU.
		\ENDFOR
		\STATE \textbf{Post-processing at central node}
		\FOR{each user $k$ at the CPU}
		\STATE Aggregate $(\mathbf R_k,\mathbf q_k)$ via \eqref{eq:CE_agg_stats} and compute $\widehat{\boldsymbol\alpha}_k$ via \eqref{eq:CE_alpha_hat}.
		\STATE Broadcast $\widehat{\boldsymbol\alpha}_k$.
		\ENDFOR
		\STATE \textbf{Output} $\{\widehat{\phi}_{k,\ell},\widehat{\alpha}_{k,\ell}\}$ and $\widehat g_{m,k}(\mathbf p_m)$.
	\end{algorithmic}
\end{algorithm}

To avoid sending the $V$-dimensional observation $\widehat{\mathbf y}_{m,k}$ from each FCA, FCA $m$ computes and forwards the following sufficient statistics for LS estimation to the CPU,
\begin{subequations}
	\label{eq:CE_suff_stats}
	\begin{align}
		\mathbf R_{m,k}
		&=
		(\mathbf A_{m,k}^{\mathrm g})^{\mathsf H}\mathbf A_{m,k}^{\mathrm g}
		\in\mathbb C^{L_k\times L_k}, \\
		\mathbf q_{m,k}
		&=
		(\mathbf A_{m,k}^{\mathrm g})^{\mathsf H}\widehat{\mathbf y}_{m,k}
		\in\mathbb C^{L_k\times 1},
	\end{align}
\end{subequations}
so that the CPU can recover the global LS solution by aggregating $\sum_m \mathbf R_{m,k}$ and $\sum_m \mathbf q_{m,k}$, while the uplink overhead scales with $L_k$ rather than $V$.

\subsubsection{Post-processing at central node}
After receiving LS sufficient statistics $\{\mathbf R_{m,k},\mathbf q_{m,k}\}_{m=1}^{M}$ from all FCAs, 
the CPU aggregates them as
\begin{align}
	\mathbf R_k
	=
	\sum_{m=1}^{M}\mathbf R_{m,k},\quad
	\mathbf q_k
	=
	\sum_{m=1}^{M}\mathbf q_{m,k}.
	\label{eq:CE_agg_stats}
\end{align}
Estimating the path gain vector, $\boldsymbol\alpha_k\in\mathbb C^{L_k\times 1}$, on fixed support $\widehat{\Omega}_k$ amounts to solving normal equation
$\mathbf R_k \boldsymbol\alpha_k = \mathbf q_k$.
To improve numerical stability when $\mathbf R_k$ is ill-conditioned, we apply diagonal loading and estimate the path-gain vector as
\begin{align}
	\widehat{\boldsymbol\alpha}_k
	=
	\big(\mathbf R_k+\epsilon_k\mathbf I_{L_k}\big)^{-1}\mathbf q_k
	\in\mathbb C^{L_k\times 1},
	\label{eq:CE_alpha_hat}
\end{align}
where $\epsilon_k>0$ is a small constant.
Note that \eqref{eq:CE_alpha_hat} only requires each antenna to transmit the sufficient statistics $(\mathbf R_{m,k},\mathbf q_{m,k})$, whose dimension scales with $L_k$ rather than the pilot block length $V$.

After obtaining $\{\widehat{\Omega}_k,\widehat{\boldsymbol\alpha}_k\}_{k=1}^{K}$ from the CPU, FCA $m$ reconstructs its effective scalar channel under any coupler position $\mathbf p_m$ as in \eqref{eq:CEc_recon}.
The reconstructed channel state information (CSI) is utilized for the subsequent coupler position optimization and digital precoding procedures described in Section~III.

The main steps of the proposed distributed channel estimation procedure are summarized in Algorithm~\ref{alg:dist_ce_fc_proxy}.
The computational complexity is dominated by pilot correlation and local proxy evaluation, and it scales as
$\mathcal O\!\big(MV\tau K + KMVG + K L_k^{3}\big)$.
Compared with the centralized channel estimation algorithm, the distributed channel estimation implementation substantially reduces the CPU-side computational burden and processing latency, especially when array size $M$ and candidate grid size $G$ are large. In terms of communication overhead, the distributed channel estimation scheme exchanges only low-dimensional LS sufficient statistics $\{\mathbf R_{m,k}, \mathbf q_{m,k}\}$ for each user $k$, whose communication load scales as $\mathcal O(MK L_k^2)$. 

\section{Simulation Results}
\begin{figure}[t!]
	\centering
	\setlength{\abovecaptionskip}{0.cm}
	\includegraphics[width=3.2in]{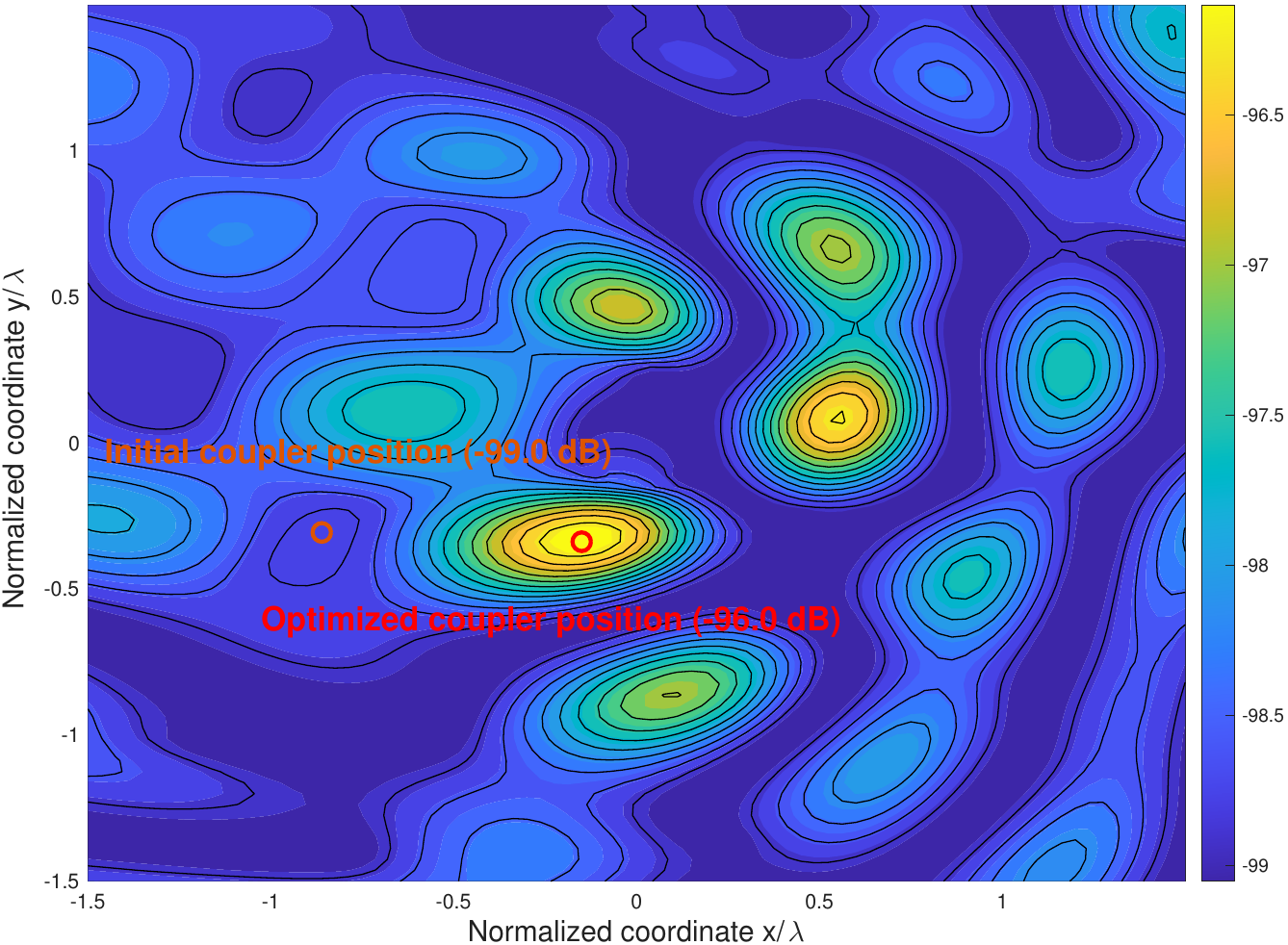}
	\caption{Channel power gain (dB) between the FCA array and
		the user with $N=1$, $K=1$, $M=5$, $P_{\max}=30$ dBm, and $L_k=15$. }
	\label{heat}
\end{figure}
In this section, numerical results are presented to evaluate the performance of the proposed distributed FCA system design. The coupler movement region associated with each active antenna is modeled as a square with side length $A$. In the simulations, both the active antennas and the passive couplers are modeled as thin straight wire dipoles. The mutual impedance matrix is computed from closed-form expressions according to \cite{balanis2016antenna}. 
The active antennas are placed with an inter element spacing of $d_y=2.2\lambda$. Each coupler is allowed to move within a square region of side length $A=2\lambda$, subject to a minimum element spacing constraint of $d_{\min}=0.15\lambda$. The carrier frequency is set to $f_c=7$~GHz. The load-impedance matrix $\mathbf X$ is diagonal with identical entries
$0.05 + j50~\Omega$.
\begin{figure}[t!]
	\centering
	\setlength{\abovecaptionskip}{0.cm}
	\includegraphics[width=2.8in]{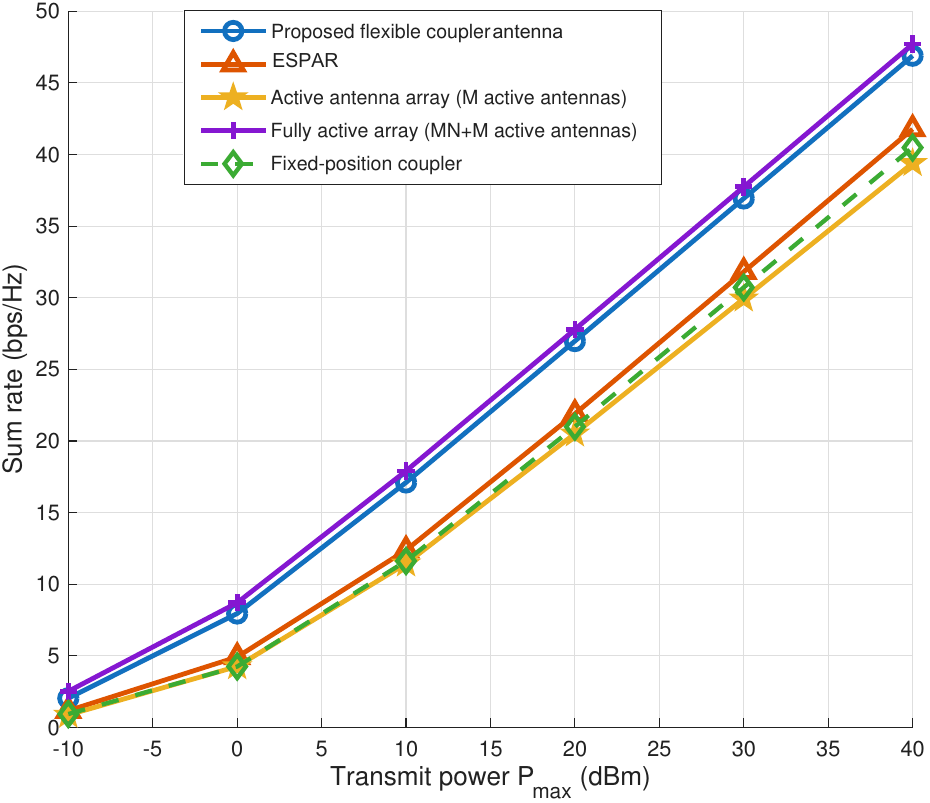}
	\caption{Achievable rate versus maximum transmit power for different schemes with $N=2$, $K=3$, $M=15$, and $L_k=15$. }
	\label{power}
	\vspace{-0.3cm}
\end{figure}

\subsection{Performance of Coupler Position Optimization}
To evaluate the performance of coupler position optimization, the proposed FCA scheme is compared with the following baseline schemes.
\begin{itemize}
	\item Active antenna array: 
	The transmitter has $M$ active antennas without any coupler elements.
	\item Fixed-position coupler: 
	The couplers are fixed at uniformly spaced positions with an inter-element spacing of $0.4\lambda$, and the load impedance matrix $\mathbf{X}$ has constant diagonal entries of $0.05 + j50~\Omega$.
	\item Electronically steerable parasitic array radiator (ESPAR)\cite{ESP}: Closed-form optimization of load impedance matrix $\mathbf X$ is used to maximize the sum rate while keeping the coupler positions fixed at the same positions as in the fixed-position coupler case.
	\item Fully active array: All $MN+M$ ports are configured as active antennas at fixed positions with half-wavelength spacing.
\end{itemize}

To demonstrate the effect of coupler position optimization on channel characteristics, Fig.~\ref{heat} shows a representative realization of the channel power gain in dB as a function of the coupler position within its feasible movement region. As can be seen, the channel power gain presents a highly irregular spatial distribution with multiple local extrema, which implies that even sub-wavelength coupler movements can significantly modify the mutual-impedance-induced currents and, consequently, the effective channel under rich multipath propagation. In particular, starting from an initial coupler location with a gain of $-99.0$~dB, the proposed algorithm steers the coupler toward a more favorable local optimum highlighted in the figure, thereby increasing the gain to $-96.0$~dB.

Fig.~\ref{power} plots the achievable sum rate versus maximum transmit power $P_{\max}$ for the proposed FCA array and several baseline schemes.
As $P_{\max}$ increases, all schemes achieve higher rates due to the improved effective signal-to-noise ratio (SNR).
It is observed that the proposed FCA array closely approaches the fully active array with $MN+M$ RF chains while using only $M$ active antennas.
This gain is achieved because coupler position optimization provides substantial mechanical beamforming and fading mitigation gains in rich multipath channels while maintaining a limited RF chain cost.
In contrast, the ESPAR scheme optimizes only the load impedances with fixed coupler locations and therefore cannot fully exploit the spatial degree of freedom (DoF) offered by position reconfiguration.
Moreover, the fixed position coupler and the active antenna array exhibit lower rates due to their fixed geometries and the lack of a reconfigurable aperture.

\begin{figure}[t!]
	\centering
	\setlength{\abovecaptionskip}{0.cm}
	\includegraphics[width=2.8in]{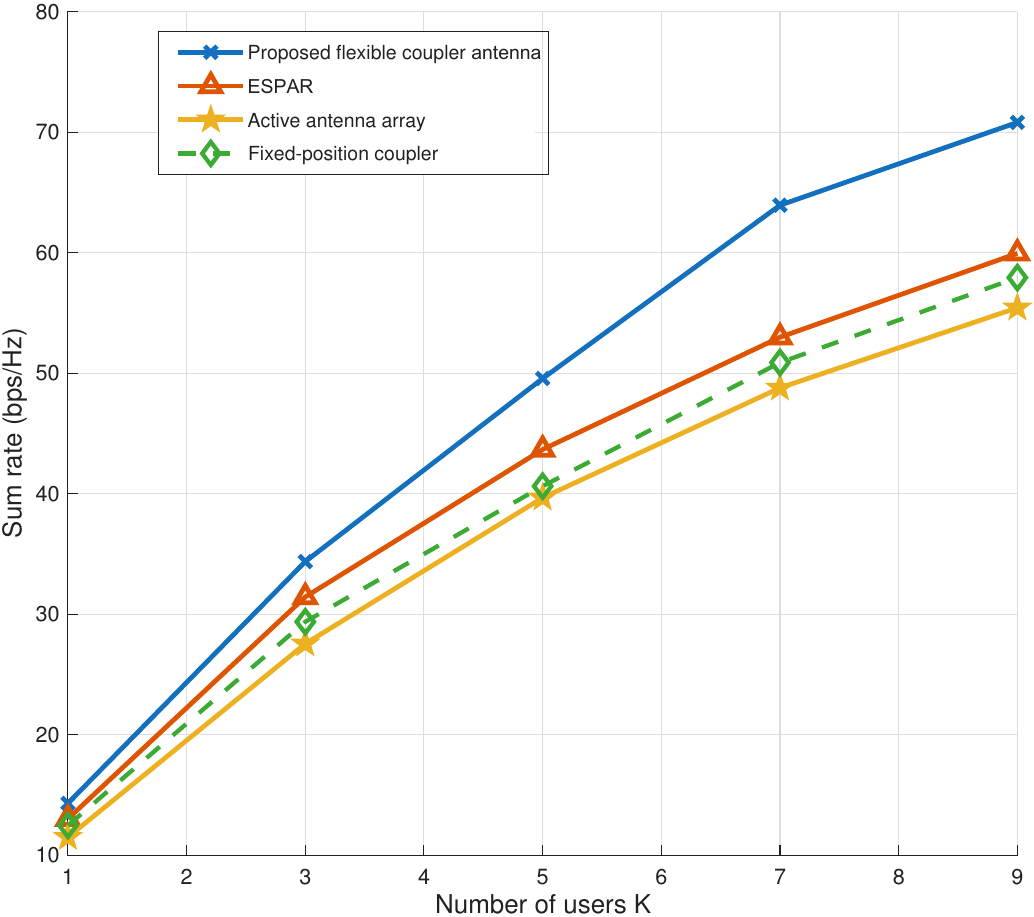}
	\caption{Achievable rate versus number of users with $N=2$, $M=10$, $P_{\max}=30$ dBm, and $L_k=15$. }
	\label{user}
	\vspace{-0.3cm}
\end{figure}
In Fig.~\ref{user}, we plot the achievable sum rate versus the number of users $K$ for the proposed FCA array and benchmark schemes.
As $K$ increases, the sum rate improves for all schemes due to the increased spatial multiplexing gain, while the growth gradually slows down since stronger multiuser interference must be mitigated under a fixed transmit power budget.
Compared with the ESPAR and fixed position coupler schemes, the proposed design consistently achieves a higher sum rate.
This performance advantage becomes more pronounced at moderate and large values of $K$, where MMSE based multiuser precoding is more sensitive to the conditioning of the effective channel matrix and thus benefits more from geometry reconfiguration, whereas the benchmark schemes with fixed geometries cannot adapt to user induced interference coupling.

\begin{figure}[t!]
	\centering
	\setlength{\abovecaptionskip}{0.cm}
	\includegraphics[width=3.39in]{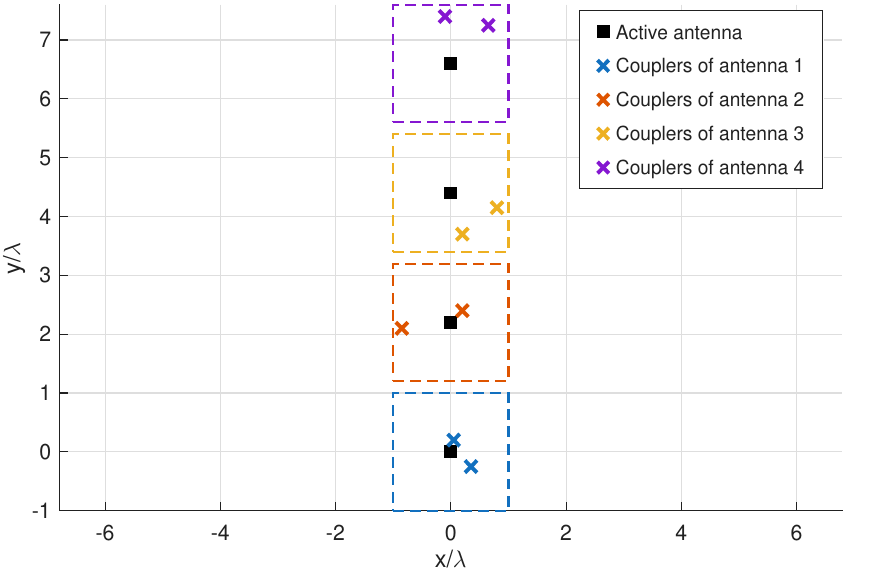}
	\caption{Optimized coupler positions obtained by the proposed algorithm with $K=2$, $M=4$, $L_k=15$, and $N=2$. }
	\label{sight}
	\vspace{-0.3cm}
\end{figure}
Fig.~\ref{sight} shows the numerically optimized coupler positions around each fixed active antenna.
It can be observed that, due to rich multipath propagation and multiuser transmission, the optimal coupler constellations differ across active antennas.
Specifically, the superposition of dominant propagation paths varies from one antenna to another, and the desired signal and inter user interference contributions are location dependent because of the coupled electromagnetic response.
As a result, the optimized coupler geometries and mechanical beamforming weights jointly enhance the effective channels of the intended users while suppressing multiuser interference through position-dependent mutual impedance.

\begin{figure}[t!] 
	\centering 
	\setlength{\abovecaptionskip}{0.cm} \includegraphics[width=2.8in]{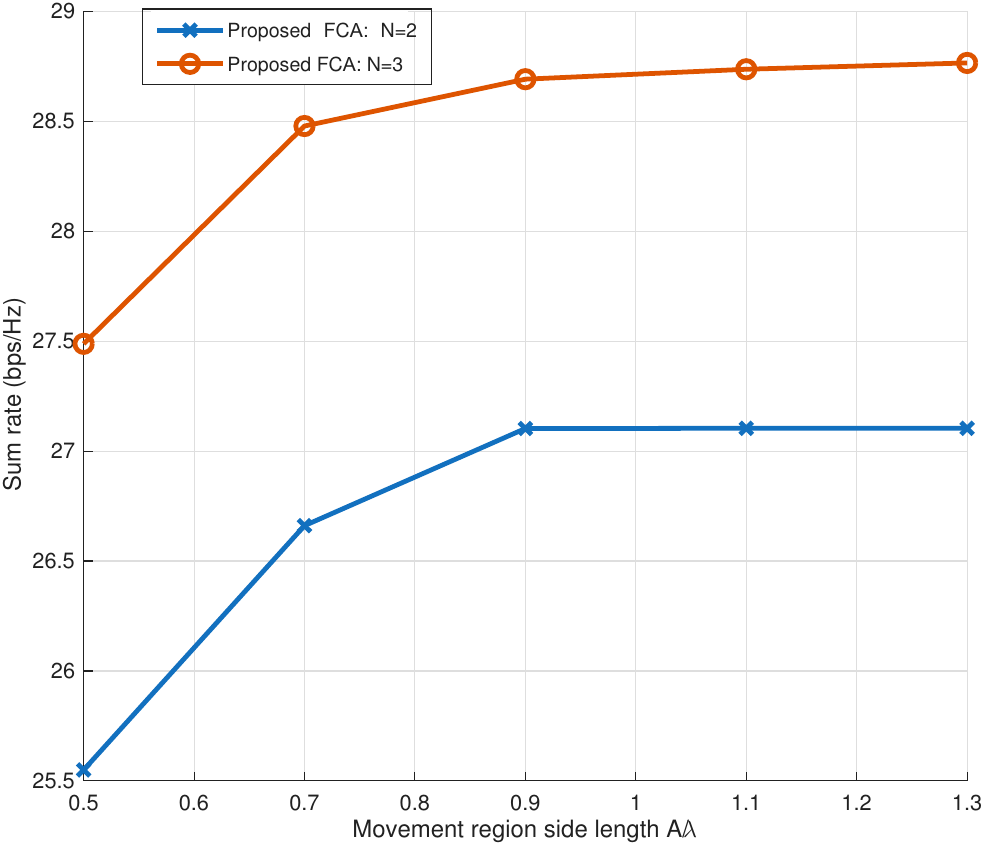} \caption{Achievable rate versus movement region size with $K=3$, $M=4$, and $L_k=15$. } \label{size} \vspace{-0.39cm} 
\end{figure}
In Fig.~\ref{size}, the achievable sum rate is illustrated as a function of the normalized movement-region side length $A/\lambda$. It can be seen that enlarging $A/\lambda$ enhances the achievable rate for both configurations, while the improvement gradually saturates as $A/\lambda$ increases. This behavior suggests that near-optimal performance can be attained with a finite movement region. For small values of $A/\lambda$, the allowable coupler positions are severely restricted, which constrains the effective aperture and reduces the diversity of the position-dependent mutual impedance, thereby weakening the resulting mechanical beamforming gain.
As $A/\lambda$ increases, position optimization becomes more effective in shaping the induced currents and mitigating unfavorable channel conditions, leading to improved sum rate. Moreover, the case with $N=3$ consistently achieves a higher sum rate than $N=2$ due to the additional passive elements and the resulting stronger reconfigurability.

\subsection{Channel estimation performance} 
Next, simulations are conducted to validate the effectiveness of the proposed distributed and centralized FCA channel estimation algorithms. The normalized mean square error (NMSE) results are obtained by averaging over all users and $500$ Monte Carlo trials.

\begin{figure}[t!]
	\centering
	\setlength{\abovecaptionskip}{0.cm}
	\includegraphics[width=2.8in]{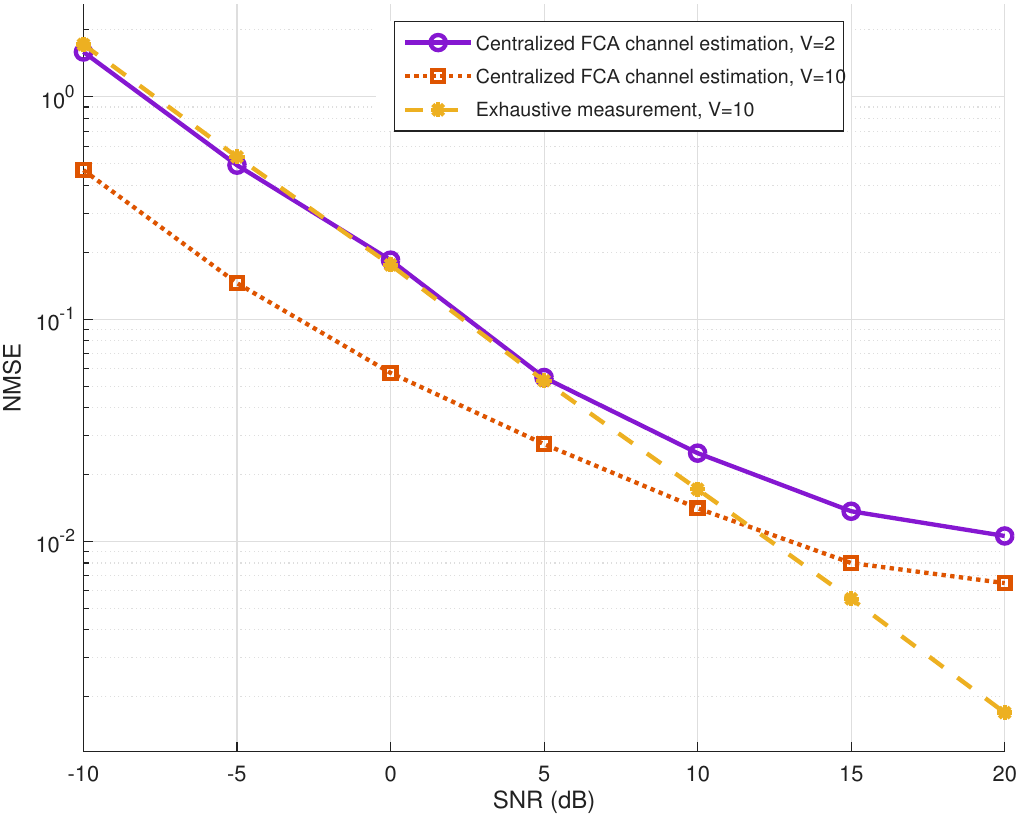}
	\caption{NMSE versus SNR for different $V$ with $K=2$, $M=8$, $N=2$,  $\tau=13$, and $L_k=3$. }
	\label{cen}
	\vspace{-0.3cm}
\end{figure}
In Fig.~\ref{cen}, we plot the NMSE versus the average received SNR for the proposed centralized FCA channel estimation with different numbers of pilot blocks $V$.
We adopt exhaustive measurement as a high-overhead benchmark scheme, where the coupler movement region is discretized into a large number ($D=400$) of candidate locations and the effective channel is sequentially measured over all candidates, resulting in on the order of $MND$ measurements per pilot block.
From Fig.~\ref{cen}, it can be seen that increasing $V$ significantly improves the NMSE performance of all schemes, especially in the low to medium SNR regime, since varying the coupler positions across blocks generates more diversified effective channel observations and enhances angular sparse reconstruction through stacked measurements.
Moreover, as observed from Fig.~\ref{cen}, exhaustive measurement becomes more competitive at high SNR due to its dense direct sampling with high computational complexity. In contrast, the proposed FCA method with a larger $V$ achieves lower NMSE at low SNR, since joint sparse reconstruction across blocks provides an inherent noise-averaging effect while requiring substantially fewer measurements.

\begin{figure}[t!]
	\centering
	\setlength{\abovecaptionskip}{0.cm}
	\includegraphics[width=2.8in]{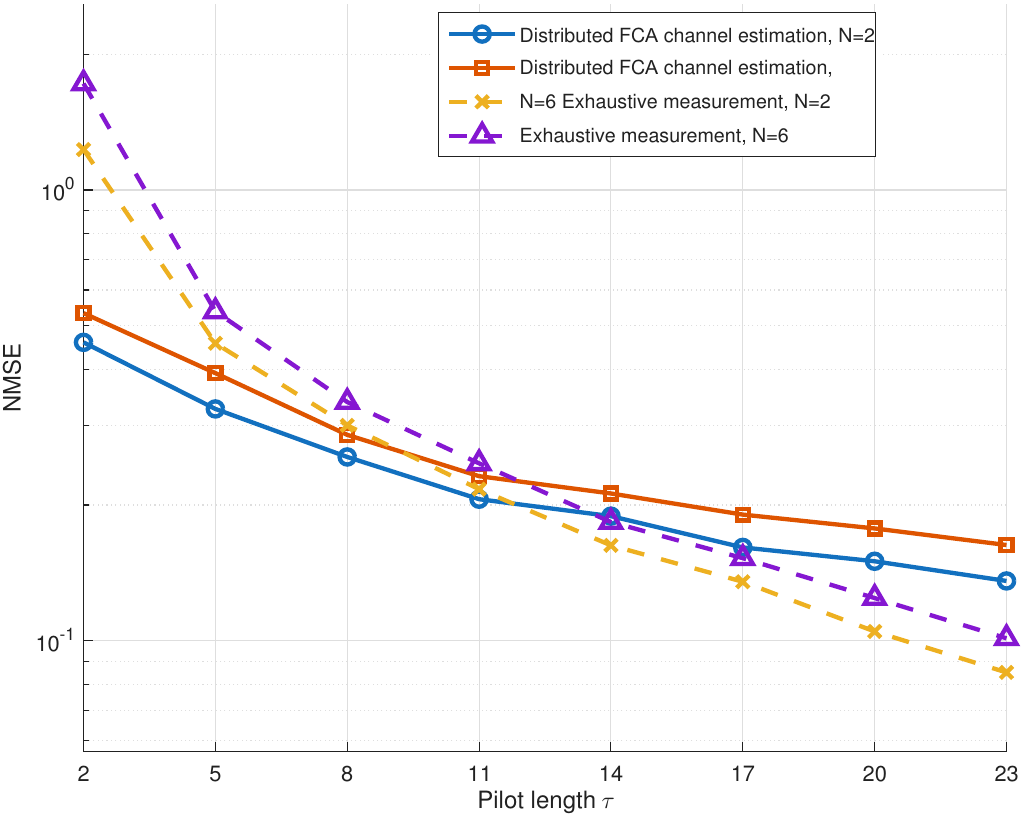}
	\caption{NMSE versus pilot length for different $N$ with $K=2$, $M=4$, SNR= $0$ dB, $\tau=13$, and $L_k=3$. }
	\label{DFCA}
	\vspace{-0.3cm}
\end{figure}
In Fig.~\ref{DFCA}, we plot the NMSE versus the pilot length $\tau$ for the proposed distributed FCA channel estimation and exhaustive measurement, under two coupler numbers $N\in\{2,6\}$.
It is observed that increasing $\tau$ improves the estimation accuracy for all schemes, since longer pilot sequences enhance the reliability of the pilot correlation outputs.
In the short pilot regime, the proposed distributed method achieves lower NMSE than exhaustive measurement for both values of $N$, as it exploits structured sparsity and performs fusion using low-dimensional sufficient statistics, which provides effective noise suppression when pilots are limited.
As $\tau$ increases, exhaustive measurement becomes more competitive due to its dense direct sampling.
Moreover, increasing $N$ leads to higher NMSE for both schemes, indicating a higher pilot requirement to maintain the same estimation accuracy.

\begin{figure}[t!]
	\centering
	\setlength{\abovecaptionskip}{0.cm}
	\includegraphics[width=2.8in]{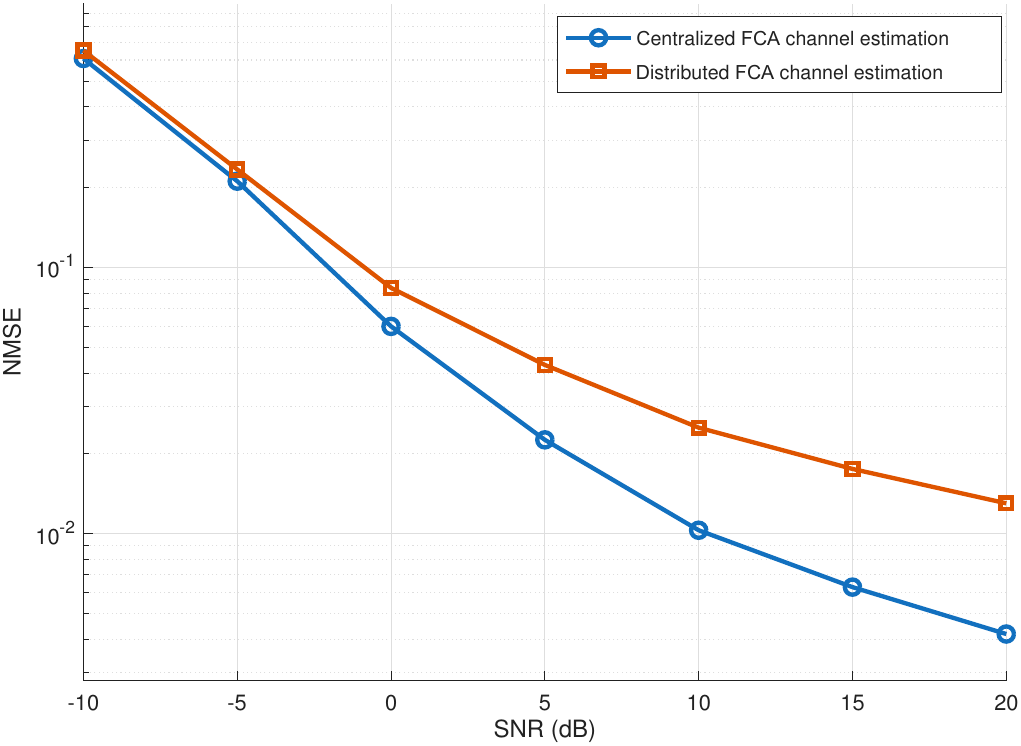}
	\caption{NMSE comparison between the centralized and distributed channel estimation schemes with \(K=2\), \(M=4\), \(N=2\), \(V=4\), \(\tau=32\), and \(L_k=10\).
	 }
	\label{discen}
		\vspace{-0.3cm}
\end{figure}
In Fig.~\ref{discen}, we compare the NMSE performance of the proposed centralized and distributed FCA channel estimation schemes versus the average receive SNR.
It is observed that both schemes achieve lower NMSE as the SNR increases, since pilot correlation becomes more reliable and stacked observations across pilot blocks provide more informative measurements for sparse reconstruction.
Moreover, the centralized scheme consistently outperforms the distributed one over the entire SNR range, as it can jointly exploit the sparsity structure across all antennas and users, whereas the distributed scheme relies on local processing at FCAs with limited information fusion.
The performance gap becomes more pronounced at medium and high SNR, where the estimation error is dominated by information loss due to decentralization rather than noise. Nevertheless, the distributed FCA channel estimation offers a favorable accuracy-complexity tradeoff, since it performs local processing and exchanges only low-dimensional sufficient statistics rather than raw pilot observations, thereby significantly reducing the communication signaling and computational burden at the CPU.

\section{Conclusions}
We have proposed a distributed FCA array that enables
mechanical beamforming through passive coupler repositioning around fixed active antennas, supported by a distributed
architecture with local LPUs and a CPU. We
have also investigated pilot-assisted centralized and distributed
channel estimation solutions for the FCA array. Simulation
results validate the performance of the proposed FCA array
and channel estimation algorithms compared with benchmark
schemes. In the proposed FCA array, reconfigurable passive
coupler positioning provides a cost-effective way to enhance
communication performance without the need to add more RF
chains. Moreover, the compact form factor and small footprint
of the FCA array make it particularly appealing for devices
with stringent size, weight, and power (SWAP) constraints.
For the future work, we will apply the proposed FCA array 
to support diverse applications and services, such as being
mounted on unmanned aerial vehicles (UAVs) with limited
payload in the low-altitude economy. Beyond communications,
we will investigate FCA-enabled target sensing to achieve a
more favorable sensing-communication tradeoff in integrated
sensing and communication (ISAC) systems. 

\bibliographystyle{IEEEtran}
\bibliography{fabs}
\end{document}